\documentclass[reprint,aps,prb,amsmath,amssymb,superscriptaddress,floatfix,citeautoscript]{revtex4-1}\usepackage[T1]{fontenc}
\usepackage[utf8]{inputenc}
\usepackage{amsmath}
\usepackage{xcolor}
\usepackage{hyperref}

\usepackage{physics,amsmath}
\usepackage{braket}
\usepackage{babel}
\usepackage{graphicx}
\usepackage{ulem}

\newcommand{\mN}{\mathcal{N}}

\usepackage{booktabs}
\usepackage{longtable}
\usepackage{array}
\newcolumntype{C}[1]{>{\centering\arraybackslash}p{#1}}

\begin{document}
\title{Size Extensive Auxiliary-Field Quantum Monte Carlo with Perturbative Coupled Cluster Trial Wavefunction} 
\author{Yichi Zhang}
\email{zhangych98@gmail.com}
\affiliation{Department of Physics, University of Colorado, Boulder, CO 80302, USA}

\author{Ankit Mahajan}
\affiliation{Department of Chemistry, Columbia University, New York, NY 10027, USA}
\author{Yann Damour}
\affiliation{Division of Chemistry and Chemical Engineering, California Institute of Technology, Pasadena, California 91125, USA}
\author{Sandeep Sharma}
\email{sanshar@gmail.com}
\affiliation{Department of Chemistry, University of Colorado, Boulder, CO 80302, USA}
\affiliation{Division of Chemistry and Chemical Engineering, California Institute of Technology, Pasadena, California 91125, USA}
\affiliation{Marcus Center for Theoretical Chemistry, Pasadena, CA 91125, USA}

\begin{abstract}
In this work, we develop a size extensive Auxiliary-Field Quantum Monte Carlo (AFQMC) approach that scales as $O(N^5)$ for local energy evaluation by treating the Coupled Cluster Singles and Doubles (CCSD) trial wavefunctions perturbatively. Comprehensive numerical examinations, spanning from main-group molecules to $3d$ transition metal complexes, demonstrate that this perturbative treatment introduces negligible bias. For small systems, our method achieves an accuracy and level of noise comparable to AFQMC with configuration interaction singles and doubles (CISD) trial wavefunctions while outperforming CCSD(T). This size extensivity offers a decisive advantage for large systems, as suggested by the ground state energies of non-interacting monomers and one-dimensional atomic chains. Finally, the numerical simulations of the uniform electron gas (UEG) provide evidence that, unlike the CCSD(T) method, our new approach does not suffer from infrared divergence in the thermodynamic limit (TDL).
\end{abstract}
\maketitle

\setlength{\tabcolsep}{10pt}

Auxiliary field quantum Monte Carlo (AFQMC) is a powerful technique for obtaining the ground state of real and model many-body quantum systems \cite{zhang1995constrained, zhang1997constrained, motta2018afqmc_rev, chen2014hubbard}. It uses imaginary-time propagation to project out all excited states from an initial state exponentially rapidly, so that at sufficiently long imaginary time, one recovers the true ground state. Performing this procedure in an unbiased way leads to an exponential increase in the Monte Carlo sampling noise with both the imaginary-time and system size, known as the sign problem \cite{feynman1979path, loh1990sign}, that renders the calculations exponentially expensive. To obtain a polynomial scaling algorithm, one modifies AFQMC by incorporating the phaseless approximation \cite{zhang2003phaseless}, which eliminates the exponentially increasing variance at the cost of a bias in the final result. The bias in the energy can be systematically reduced by improving the quality of the trial state used to impose the constraint. The AFQMC energy is given by
\begin{equation}
    E = \frac{\braket{\psi_T|\hat{H}|\Psi_\text{AF}}}{\braket{\psi_T|\Psi_\text{AF}}} ,\label{eq:e_afqmc}
\end{equation}
where $\ket{\psi_T}$ is the trial state, $\ket{\Psi_\text{AF}}$ is the AFQMC wavefunction, and $\hat{H}$ is the Hamiltonian. In the limit of an exact trial state, the phaseless-AFQMC (ph-AFQMC) energy approaches the exact ground state energy with vanishing sampling noise, a property referred to as the zero variance principle. Note that in this limit the AFQMC wavefunction is not guaranteed to be exact, as demonstrated in recent work by Li and coworkers \cite{li2025fokker}.

This has led several groups to formulate efficient AFQMC algorithms that exploit more accurate and sophisticated trial states, including matrix product states \cite{wouters2014mps_afqmc}, selected configuration interaction wavefunctions \cite{mahajan2022sci_afqmc}, variational AFQMC (VAFQMC) \cite{xiao2025sto}, truncated configuration interaction wavefunctions \cite{mahajan2025afqmc_ci, kjonstad2025systematic}, etc. A recent study by our group proposed the use of truncated configuration interaction wavefunctions including singles and doubles excitations as a trial state (AFQMC/CISD) \cite{mahajan2025afqmc_ci}, which led to a method that scales as $O(N^5)$ for the local energy evaluation (once the trial state is generated), where $N$ denotes the system size. Through a comprehensive set of benchmarks, the authors demonstrated that AFQMC/CISD is often more accurate than CCSD(T) - widely regarded as the gold standard in quantum chemistry that scales as $O(N^7)$\cite{paldus1999coupled, bartlett2007coupled} - despite a lower computational scaling. 

A key remaining shortcoming of AFQMC/CISD is that the energies are not size-extensive: the sum of all non-interacting subsystems does not yield the same energy as the total system. The size-extensivity errors in AFQMC/CISD are smaller than those of the underlying CISD trial state itself, which makes the method useful for small to medium-sized molecules, but will likely preclude its applications to large molecules and material science, where the proper scaling of the correlation energy (and all extensive properties in general) with system size is important\cite{lyakh2012multireference, nooijen2005reflections}. The lack of size-extensivity also poses a fundamental obstacle to the development of cluster-in-molecule (CIM) style linear-scaling methods\cite{li2002linear,  rolik2011general, rolik2013efficient, nagy2019approaching}, which have been recently demonstrated for AFQMC/HF \cite{kurian2023linear}. The goal of this article is to present some modified algorithms that restore size-extensivity without deteriorating the accuracy of AFQMC/CISD for small systems.  

The remainder of the article is organized as follows. In Sec.~\ref{sec:background}, we introduce the theoretical background of phaseless AFQMC. In Sec.~\ref {sec:size}, we discuss the size-extensive error in AFQMC/CISD and present the size-extensive alternatives developed in this work. In Sec.~\ref{sec:numerical}, we provide numerical benchmarks that support the size-extensivity and high-accuracy of the new approach. In Sec.~\ref{sec:conclusion}, we summarize the advantages and limitations of the proposed methods and outline directions for future improvement. More details and data supporting this work are presented in the supporting information. 

\section{Background}
\label{sec:background}

In the AFQMC algorithm, one works with the second-quantized Hamiltonian in the form
\begin{equation}
\begin{split}
    \hat{H} =& H_0 + \hat{H}_1 + \hat{H}_2  \\
    =& H_0 + \sum_{pq} (h_{pq}-\frac{1}{2}\sum_r v_{prrq}) \hat{a}_p^\dagger \hat{a}_q  \\
    & +\frac{1}{2}\sum_{pqrs} v_{pqrs} \hat{a}_p^\dagger \hat{a}_r \hat{a}_q^\dagger \hat{a}_s
\end{split}
\end{equation}
where the $H_0$, $\hat{H}_1$, $\hat{H}_2$  are the core energy, one-body term, and the two-body term, while $h_{pq}$ and $v_{pqrs}$ are the one- and two-electron integrals, respectively. Before proceeding, we perform a Cholesky decomposition on the two-electron integrals 
\begin{equation}
  v_{pqrs} = \sum_\gamma L_{\gamma pr} L_{\gamma qs}
\end{equation}
which allows us to write the two-body operator as a sum of squares of one-body operators
\begin{equation}
\begin{split}
    \hat{H} =& H_0 + \hat{H}_1 - \frac{1}{2} \sum_\gamma \hat{v}_\gamma^2
\end{split}
\end{equation}
where $\hat{v}_\gamma =  i\sum_{pq} L_{\gamma pq} \hat{a}_p^\dagger \hat{a}_q$. It is often advantageous to perform a so-called mean-field subtraction, which changes the various terms of the Hamiltonian without modifying the overall Hamiltonian as follows 
\begin{equation}
\begin{split}
    H_0 \rightarrow &  H_0 + \frac{1}{2}\sum_\gamma m_\gamma^2 \\
    \hat{H}_1 \rightarrow & \hat{H}_1 - \sum_\gamma m_{\gamma} \hat{v}_{\gamma}\\
    \hat{v}_\gamma\rightarrow & \frac{1}{2} \sum_\gamma (\hat{v}_\gamma- m_\gamma)^2
\end{split}
\end{equation}
where $m_\gamma$ is a constant chosen as
\begin{equation}
\begin{split}
    m_\gamma = i\sum_{pq} L_{\gamma pq} \frac{\langle\psi_T|\hat{a}_p^\dagger \hat{a}_q|\psi_T\rangle}{\langle\psi_T|\psi_T\rangle}
\end{split}
\end{equation}
This modification is known to reduce the noise in the stochastic simulation. $\psi_T$ is a trial state and will play an important role in the phaseless approximation. After an initial state $|\phi(0)\rangle$ is chosen, an imaginary time propagation is performed to give the state ($|\phi(\tau)\rangle$) at a subsequent imaginary time ($\tau$)
\begin{equation}
\begin{split}
    |\phi(\tau)\rangle = & \exp(-\tau (\hat{H}-E_0))|\phi(0)\rangle \\ 
    = & e^{-\tau(H_0-E_0)} \times \\
    & \int \frac{d\mathbf{x}}{(2 \pi)^{N_\gamma/2}} e^{\frac{-\mathbf{x}^2}{2}} e^{-\frac{\delta t \hat{H}_1}{2}}e^{\sqrt{\delta t}\mathbf{x}\cdot \mathbf{\hat{v}}}e^{-\frac{\delta t\hat{H}_1}{2}}|\phi(0)\rangle \\
    & + O(\delta t^2)
\end{split}
\end{equation}
where we have used the Trotter decomposition and the Hubbard-Stratonovich transformation to go from the first to the second step. $E_0$ is the ground state energy estimation that is dynamically updated to stabilize the walker population. The algorithm can be viewed as an iterative imaginary time propagation where the wavefunction at time $n\delta t$ is obtained from that at time $(n-1)\delta t$ using
\begin{equation}
\begin{split}
    |\Psi^{(n)}\rangle =& \int d\mathbf{x}_n p(\mathbf{x}_n) \hat{B}(\mathbf{x}_n) |\Psi^{(n-1)}\rangle 
\end{split}
\end{equation}
where the operator $\hat{B}(\mathbf{x})$ is a product of exponentials of one-body operators which can be applied exactly using the Thouless theorem if the state $|\Psi_{n-1}\rangle$ is a linear combination of single Slater determinants. This high-dimensional integral is calculated using Monte Carlo sampling, and the (free projection) AFQMC wavefunction at step $n$ is given by
\begin{equation}
\begin{split}
 |\Psi^{(n)}\rangle = \sum_i w_i^{(n)} |\phi_i^{(n)}\rangle \label{eq:sample}
\end{split}
\end{equation}
where $w_i^{(n)}$ and $|\phi_i^{(n)}\rangle $ are respectively the weights and walkers sampled at step $n$. 
So far, no approximations have been made, and Eq.~\ref{eq:sample} gives an unbiased sample of the true ground state wavefunction after a sufficiently large number of steps $n$ have elapsed.

However, the method suffers from the sign problem except in cases of Hamiltonians having special symmetries. To overcome the sign problem, one introduces importance sampling and phaseless approximation. Importance sampling amounts to performing a similarity transformation on the Hamiltonian, which does not change the eigenvalue but modifies the eigenvector in a way that makes it more efficient to sample. Importance sampling is representation dependent and, in AFQMC, it amounts to updating the walkers and weights as follows
\begin{equation}
\begin{split}
|\phi_i^{(n)}\rangle \leftarrow & \hat{B}^{(n-1)}(\mathbf{x})|\phi_i^{(n-1)}\rangle \\
w_i^{(n)} \leftarrow & w_i^{(n-1)}\frac{\langle\psi_T|\hat{B}^{(n-1)}(\mathbf{x})|\phi_i^{(n-1)}\rangle}{\langle\psi_T|\phi_i^{(n-1)}\rangle} \mN^{(n-1)}(\mathbf{x})
\label{eq:impsample}   
\end{split}
\end{equation}
where the modified propagator $\hat{B}(\mathbf{x})$ is given by
\begin{equation}
\begin{split}
 \hat{B}^{(n)}(\mathbf{x}) = & e^{-\frac{\delta t \hat{H}_1}{2}}e^{\sqrt{\delta}(\mathbf{x}-\mathbf{f}^{(n)})\cdot \mathbf{\hat{v}}} e^{-\frac{\delta t \hat{H}_1}{2}} \label{eq:prop}\\
 \mN^{(n)}(\mathbf{x})= & e^{-\frac{\mathbf{f}^{(n)}\cdot \mathbf{f}^{(n)}}{2}+\mathbf{x}\cdot\mathbf{f}^{(n)}}
\end{split}
\end{equation}
where $\mathbf{f}$ is called the force biases and is given as 
\begin{equation}
\begin{split}
    f_\gamma^{(n)} = - \sqrt{\delta t } \frac{\langle\psi_T|\hat{v}_\gamma|\phi^{(n)}\rangle} {\langle\psi_T|\phi^{(n)}\rangle} 
\end{split}
\end{equation}

Contrary to popular belief, using importance sampling introduces a bias in the AFQMC wavefunction even if the trial state is the exact ground state, 
as shown recently in Ref.~\cite{li2025fokker}. Within the importance sampling approximation, the state sampled by the AFQMC algorithm is given by
\begin{equation}
\begin{split}
 |\Psi^{(n)}\rangle = \sum_i \frac{w_i^{(n)}}{\langle\psi_T|\phi_i^{(n)}\rangle} |\phi_i^{(n)}\rangle    
\end{split}
\end{equation}
The ground state energy is evaluated as the projected energy
\begin{align}
    E^{(n)} =& \frac{\langle\psi_T|\hat{H}|\Psi^{(n)}\rangle}{\langle\psi_T|\Psi^{(n)}\rangle} \nonumber \\
    =& \frac{\sum_i w_i^{(n)} E_L(\phi_i^{(n)})}{\sum_i w_i^{(n)}} \label{eq:locSample}
\end{align}
where $E_L(\phi_i^{(n)})$ is the local energy defined for walker $i$ at step $n$:
\begin{align}
    E_L(\phi_i^{(n)})=& \frac{\langle\psi_T|\hat{H}|\phi_i^{(n)}\rangle}{\langle\psi_T|\phi_i^{(n)}\rangle} \label{eq:locE}
\end{align}
Although importance sampling introduces bias in the AFQMC wavefunction, the local energy in (\ref{eq:locE}) is exact if the trial state is exact due to the zero-variance principle. However, the importance sampling by itself does not eliminate the sign problem because the weights $w_i^{(n)}$ in the denominator of (\ref{eq:locSample}) can take any value in the complex plane and their sum can come arbitrarily close to zero. This is overcome by introducing the phaseless approximation which breaks gauge invariance in the complex plane as follows
\begin{equation}
\begin{split}
w_i^{(n)}  =& \left| w_i^{(n)}\right| \times \max(0, \cos(\theta)) \\
\theta =& \arg\left[ \frac{\langle\psi_T|\hat{B}(\mathbf{x})|\phi_i^{(n-1)}\rangle}{\langle\psi_T|\phi_i^{(n-1)}\rangle} \right] 
\label{eq:impWeight}
\end{split}
\end{equation}
and ensures that the weights are positive real numbers. This eliminates the sign problem at the cost of introducing a bias in the energy for approximate trial wavefunctions. 

The algorithm thus proceeds in two steps:
\begin{enumerate}
    \item We perform propagation by acting the operator $\hat{B}(\mathbf{x})$, shown in (\ref{eq:prop}), to walkers using the auxiliary fields $\mathbf{x}$ sampled from the Gaussian distribution. The update to the walker is followed by an update to the weight as given in (\ref{eq:impsample}) and (\ref{eq:impWeight}).
    \item After a sufficiently large number of propagation steps are performed, we sample the energy by using eq.(\ref{eq:locSample}). The energy samples are averages over several steps to obtain the total energy and standard deviation. 
\end{enumerate}

\section{Size-extensivity} 
\label{sec:size}

Following the discussion by Lee et al. we show that AFQMC energy is size-extensive in the small time-step limit, provided the trial wavefunction is a product state\cite{lee2022twenty}. Consider a system composed of two non-interacting subsystems A and B. If the trial state has a product structure, that is $|\psi_T\rangle=|\psi_{T,A}\rangle|\psi_{T,B}\rangle$, then the propagator likewise factorizes as $\hat{B}(\mathbf{x}) = \hat{B}_A(\mathbf{x}_A)\hat{B}_B(\mathbf{x}_B)$, where the propagators of A and B commute with each other. This implies that if our walker is initialized as a product state with the correct number of electrons on the fragments, it remains a product state in the absence of the phaseless approximation. 

The phaseless approximation, however, introduces a non-size extensivity error because it includes the cosine term in (\ref{eq:impWeight}) which is not product separable
\begin{equation}
\begin{split}
    \cos(\theta_A + \theta_B) = \cos(\theta_A)\cos(\theta_B) - \sin(\theta_A)\sin(\theta_B)
\end{split}
\end{equation}
where $\theta_{A/B} \propto \sqrt{\delta t} x_{A/B} + \delta t$, with the two contributions arising from the two-body and one-body terms, respectively. The order of the resulting size-extensivity error can be estimated as
\begin{equation}
\begin{split}
    &\langle \sin(\theta_A)\sin(\theta_B)\rangle \\
    & = \langle x_A x_B \delta t \rangle_{p_Ap_B} \\
    & \quad +\langle (x_A + x_A^2 x_B + x_B + x_B^2 x_A) \delta t^{3/2}\rangle_{p_Ap_B} \\
    & \quad +\langle \delta t^2\rangle + O(\delta t^4) \label{eq:hf_size_error}
\end{split}
\end{equation}
where $p_A$ and $p_B$ are Gaussian distributions of $x_A$ and $x_B$, respectively. Note that the first two terms average to zero because either $x_A$ or $x_B$ appears with odd powers, and integrating them against a Gaussian gives zero. The size-extensivity error per time step is therefore $O(\delta t^2)$, and for the entire simulation it scales as $O(\delta t)$, which is the same as the phaseless AFQMC time-step error. Throughout this work we use $\delta t = 0.005$~a.u., which Lee and coworkers showed to be sufficient to suppress this error\cite{lee2022twenty}.

If the AFQMC wavefunction being sampled and the trial state have a multiplicative form, then the AFQMC energy as given in (\ref{eq:locSample}) will be size-extensive. This no longer holds in general for multideterminantal wavefunctions, including the configuration interaction singles and doubles that some of the authors have recently introduced~\cite{mahajan2025afqmc_ci}.

\subsection{Perturbative AFQMC/CCSD Energy}

Because the coupled-cluster singles and doubles (CCSD) wavefunction is product-separable, one would expect AFQMC energies to be size-extensive when it is used as a trial state. However, a direct evaluation of the energy, overlap, and force bias with a CCSD trial wavefunction is computationally expensive. In prior work~\cite{mahajan2025afqmc_ci}, we showed that a CISD-like wavefunction — obtained by applying a projector $\hat{P}_{D}$ onto the space of single and double excitations (i.e., $\hat{P}_{D} e^{\hat{T}} \ket{\psi_0}$) — yields highly accurate energies. Nevertheless, the introduction of this projector breaks product separability, meaning size-extensivity is no longer guaranteed.

Here, we introduce a perturbative expression for the AFQMC/CCSD energy that recovers size-extensivity without requiring an exact application of the CCSD trial state. As we will demonstrate in the results section, this perturbative approach yields energies of comparable accuracy to the AFQMC/CISD method~\cite{mahajan2025afqmc_ci}. To simplify the derivation, we work with the full cluster operator without decomposing it into singles and doubles; the formula will be specialized to CCSD later. In the energy expression for AFQMC with a coupled-cluster trial, we introduce a parameter $\lambda$ into both the numerator ($N(\lambda)$) and the denominator ($D(\lambda)$) to track the order of the cluster operator:

\begin{equation}
\begin{split}
    &N(\lambda) = \bra{\psi_0} e^{\lambda \hat{T}^\dagger} \hat{H} \ket{\Psi_\text{AF}} \\
    &D(\lambda) = \bra{\psi_0} e^{\lambda \hat{T}^\dagger} \ket{\Psi_\text{AF}} \\
    &E = \left. \frac{N(\lambda)}{D(\lambda)} \right|_{\lambda=1} = \left. \sum_{n} \lambda^n E^{[n]} \right|_{\lambda=1} \label{eq:pt}
\end{split}
\end{equation}

By Taylor expanding $N(\lambda)$ and $D^{-1}(\lambda)$ with respect to $\lambda$, we obtain the perturbative energy $E^{[n]}$ at each order. (We use square brackets to denote the perturbation order, to distinguish them from the parentheses used in the previous section for imaginary time steps.) The expressions for $E^{[n]}$ up to the second order are:
\begin{equation}
\begin{split}
E^{[0]} & =\frac{\braket{\psi_{0}|\hat{H}|\Psi_\text{AF}}}{\braket{\psi_{0}|\Psi_\text{AF}}} \\
E^{[1]} & =\frac{\braket{\psi_{0}|\hat{T}^\dagger\hat{H}|\Psi_\text{AF}}}{\braket{\psi_{0}|\Psi_\text{AF}}}-E^{[0]}\frac{\braket{\psi_{0}|\hat{T}^\dagger|\Psi_\text{AF}}}{\braket{\psi_{0}|\Psi_\text{AF}}} \\
E^{[2]} & =\frac{1}{2}\frac{\braket{\psi_{0}|\hat{T}^\dagger{}^{2}\hat{H}|\Psi_\text{AF}}}{\braket{\psi_{0}|\Psi_\text{AF}}}-E^{[0]}\frac{\braket{\psi_{0}|\hat{T}^\dagger{}^{2}|\Psi_\text{AF}}}{\braket{\psi_{0}|\Psi_\text{AF}}} \\ & \quad- 2E^{[1]}\frac{\braket{\psi_{0}|\hat{T}^\dagger|\Psi_\text{AF}}}{\braket{\psi_{0}|\Psi_\text{AF}}} \label{eq:locPT}
\end{split}
\end{equation}

If we perform AFQMC calculations using the Hartree-Fock (HF) wavefunction as the guiding state (which is used to evaluate the force bias and impose the phaseless constraint), the AFQMC state $\Psi_{AF}$ also inherently possesses a product structure in the limit of small time steps, according to our previous argument regarding Eq.~\ref{eq:hf_size_error}. Considering $N$ non-interacting subsystems, each denoted by a superscript ``$a$'', the cluster operator, the Hamiltonian, the HF state, and the AFQMC state can be expressed as:

\begin{equation}
\begin{split}
 \hat{T}=& \sum_{a=1}^N \hat{T}^a \qquad\hat{H}=\sum_{a=1}^N \hat{H}^a \\
 |\psi_0\rangle =& \bigotimes_{a=1}^N |\psi_0^a\rangle \qquad |\Psi_\text{AF}\rangle = \bigotimes_{a=1}^N |\Psi_\text{AF}^a\rangle
\end{split}
\end{equation}
The cluster operator $\hat{T}$ is additive separable in the Coupled-Cluster theory. It is straightforward to show that $E^{[0]}$ is extensive:
\begin{equation}
\begin{split}
    E^{[0]} & = \sum_{a=1}^N E^{[0],a}\\
    E^{[0],a} &= \frac{\braket{\psi^a_{0}|\hat{H}|\Psi^a_\text{AF}}}{\braket{\psi^a_{0}|\Psi^a_\text{AF}}} \\
\end{split}
\end{equation}
The extensivity of $E^{[1]}$ follows:

\begin{equation}
\begin{split}
E^{[1]} = & \sum_{a=1}^N E^{[1],a}\\
E^{[1],a}= & \frac{\braket{\psi_0^a|\hat{T}^a{}^\dagger\hat{H}^a |\Psi_\text{AF}^a}}{\braket{\psi_{0}^a|\Psi_\text{AF}^a}}-E^{[0],a}\frac{\braket{\psi_{0}^a|\hat{T}^a{}^\dagger|\Psi_\text{AF}^a}}{\braket{\psi_{0}^a|\Psi_\text{AF}^a}}
\end{split}
\end{equation}
Note that the two terms in the $E^{[1]}$, namely $\frac{\braket{\psi_{0}|\hat{T}^\dagger\hat{H}|\Psi_\text{AF}}}{\braket{\psi_{0}|\Psi_\text{AF}}}$ and $E^{[0]}\frac{\braket{\psi_{0}|\hat{H}|\Psi_\text{AF}}}{\braket{\psi_{0}|\Psi_\text{AF}}}$ are individually non-size-extensive and contain terms that scales quadratically with the number of fragments. However, these non-size-extensive contributions exactly cancel out when we take their difference. A similar proof can be carried out for $E^{[2]}$ as well.

A more general argument of size-extensivity can be established using the order parameter $\lambda$. For two non-interacting fragments $a$ and $b$, we have $e^{\lambda \hat{T}^{\dagger}} = e^{\lambda \hat{T}^{a,\dagger}} e^{\lambda \hat{T}^{b,\dagger}}$, which naturally leads to $E(\lambda) = E^a(\lambda) + E^b(\lambda)$. By expanding both sides as a power series in $\lambda$ and equating the terms at each order $\lambda^n$, we obtain $E^{[n]} = E^{[n],a} + E^{[n],b}$, demonstrating that size-extensivity is preserved at all perturbation orders.

\subsection{Three levels of approximations}
Having outlined the expression for obtaining size-extensive energies, we will introduce three different variants: (1) both the $e^{\hat{T}_1}$ and $e^{\hat{T}_2}$ cluster operators are treated perturbatively to the first order, (2) $e^{\hat{T}_1}$ is treated exactly to all orders by Thouless theorem and $e^{\hat{T}_2}$ is treated perturbative to the first order and (3) in which both $e^{\hat{T}_1}$ and $e^{\hat{T}_2}$ are treated to all order, $e^{\hat{T}_2}-1-\hat{T}_2$ is stochastically sampled by Hubbard-Stratonovich decomposition. The order of perturbation is chosen to ensure that the cost of the local energy evaluation has the same scaling as in AFQMC/CISD.

\subsubsection{$e^{\hat{T}_1}$ and $e^{\hat{T}_2}$ perturbatively}
In the first method, we treat both singles and doubles perturbatively. If $\hat{T}=\hat{T}_1+\hat{T}_2$ then we get
\begin{equation}
\begin{split}
    E^{[0]} =& \frac{\braket{\psi_0|\hat{H}|\Psi_\text{AF}}}{\braket{\psi_0|\Psi_\text{AF}}} \\
    =& \frac{\sum_i w_i E_L^{0,1}(\phi_i)}{\sum_i w_i} \\
    E^{[1]} =& \frac{\braket{\psi_0|\hat{T}^\dagger \hat{H}|\psi_\text{AF}}}{\braket{\psi_0|\psi_\text{AF}}} - E^{[0]}\frac{\braket{\psi_0|\hat{T}^\dagger|\psi_\text{AF}}}{\braket{\psi_0|\psi_\text{AF}}} \\
    =& \frac{\sum_iw_iE_L^{1,1}(\phi_i)}{\sum_iw_i} - E^{[0]}\frac{\sum_iw_iE_L^{1,0}(\phi_i)}{\sum_iw_i} \label{eq:ept1}
\end{split}
\end{equation}
where we have introduced the notation 
\begin{equation}
    E_L^{m,n}(\phi_i)  =\frac{\braket{\psi_0|(\hat{T}^\dagger)^m \hat{H}^n|\phi_i }}{\braket{\psi_0|\phi_i}}
\end{equation}

In this formulation, the guiding wavefunction is the Hartree-Fock state. We refer to this as AFQMC/ptCCSD (or PT for short).

\subsubsection{$e^{\hat{T}_1}$ non-perturbatively and $e^{\hat{T}_2}$ perturbatively}
In the second approach, we use the fact that $e^{\hat{T}_1}$ can be applied to the mean field exactly using the Thouless theorem. This allows us to write an energy expression where only the $\hat{T}_2$ are treated perturbatively and we introduce the wavefunction $|\tilde{\psi}_0\rangle = e^{\hat{T}_1}|\psi_0\rangle$. The energy becomes
\begin{equation}
    E = \tilde{E}^{[0]} + \tilde{E}^{[1]}
\end{equation}
where the zeroth- and first-order energy expressions are

\begin{equation}
\begin{split}
    \tilde{E}^{[0]} &= \frac{\braket{\tilde{\psi}_0|\hat{H}|\Psi_\text{AF}}}{\braket{\tilde{\psi}_0|\Psi_\text{AF}}} \\
    \tilde{E}^{[1]} &= \frac{\braket{\tilde{\psi}_0|\hat{T}_2^\dagger \hat{H}|\Psi_\text{AF}}}{\braket{\tilde{\psi}_0|\Psi_\text{AF}}} - \tilde{E}^{[0]}\frac{\braket{\tilde{\psi}_0|\hat{T}_2^\dagger|\Psi_\text{AF}}}{\braket{\tilde{\psi}_0|\Psi_\text{AF}}} \\
\end{split}
\end{equation}
In this formulation, it seems natural to use $\ket{\tilde{\psi_0}}$ as the guiding wavefunction; however, we find that doing so leads to larger errors in the total energy. To avoid the deterioration in the quality of results, we use $\ket{\tilde{\psi_0}}$ as the trial state ($|\psi_T\rangle$) for projected energy evaluation, and the Hartree-Fock state $\ket{\psi_0}$ as the guiding wavefunction ($|\psi_G\rangle$). The AFQMC energy then becomes 
\begin{equation}
\begin{split}
    \frac{\braket{\psi_T|\hat{H}|\Psi_\text{AF}}}{\braket{\psi_T|\Psi_\text{AF}}} =& \frac{\braket{\psi_T|\hat{H}|\Psi_\text{AF}}}{\braket{\psi_G|\Psi_\text{AF}}}/\frac{\braket{\psi_T|\Psi_\text{AF}}}{\braket{\psi_G|\Psi_\text{AF}}} \\
    =& \frac{\sum_i w_i\frac{\braket{\psi_T|\hat{H}|\phi_i}}{\braket{\psi_G|\phi_i}}}{\sum_i w_i} / \frac{\sum_i w_i\frac{\braket{\psi_T|\phi_i}}{\braket{\psi_G|\phi_i}}}{\sum_i w_i} \\
    =& \frac{\sum_i w'_iE_L(\phi_i)}{\sum_i w'_i} \label{eq:guide}
\end{split}
\end{equation}
where $w'_i = w_i\frac{\braket{\psi_T|\phi_i}}{\braket{\psi_G|\phi_i}}$ and $E_L(\phi_i) = \frac{\braket{\psi_T|\Omega|\phi_i}}{\braket{\psi_T|\phi_i}}$. One can readily extend this result to (\ref{eq:locPT}) and obtain
\begin{equation}
\begin{split}
    \tilde{E}^{[0]} =& \frac{\sum_i \tilde{w}_i \tilde{E}_L^{0,1}(\phi_i)}{\sum_i \tilde{w}_i} \\
    \tilde{E}^{[1]} =& \frac{\sum_i \tilde{w}_i \tilde{E}_L^{1,1}(\phi_i)}{\sum_i \tilde{w}_i} - \tilde{E}^{[0]} \frac{\sum_i \tilde{w}_i \tilde{E}_L^{1,0}(\phi_i)}{\sum_i \tilde{w}_i} \label{eq:pt2}
\end{split}
\end{equation}
where we have introduced the notation
\begin{equation}
\begin{split}
    \tilde{E}_L^{m,n}(\phi_i) =& \frac{\langle\tilde{\psi}_0|(\hat{T}^\dagger_2)^m \hat{H}^n|\phi_i\rangle }{\langle\tilde{\psi}_0|\phi_i\rangle} \\
    \tilde{w}_i =& w_i \frac{\braket{\tilde{\psi}_0|\phi_i}}{\braket{\psi_0|\phi_i}} \label{eq:tilde}
\end{split}
\end{equation}

We refer to this as AFQMC/pt2CCSD (or PT2 for short). As we will see in the results section, AFQMC/pt2CCSD is more accurate than AFQMC/ptCCSD.

\subsubsection{$e^{\hat{T}_1}$ and $e^{\hat{T}_2}$ non-perturbatively}
Finally, we introduce a formulation in which both the $\hat{T}_1$ and $\hat{T}_2$ are treated to all orders. To proceed, we first absorb $e^{\hat{T_1}}$ into the Hartree-Fock reference and obtain $\ket{\tilde{\psi}_0}$ as before. Then, one performs an eigenvalue decomposition on the $\hat{T_2}$ amplitude
\begin{equation}
    t_{iajb} = Q^T_{(ia),g} \Sigma_g Q_{g,(jb)} = \tau_{gia}\tau_{gjb}
\end{equation}
where $\tau_{gia} = \sqrt{\Sigma_g} Q_{gia}$. This is very similar to the Cholesky decomposition that AFQMC uses, but the CCSD amplitudes are not positive semi-definite, 
so the three-index tensors $\tau$ are generally complex. Like Cholesky decomposition, the convenience of such decomposition is that it allows us to represent the exponential of the $e^{\hat{T}_2}$ operator stochastically by the HS-transformation:
\begin{equation}
\begin{split}
    e^{\hat{T}_2} \ket{\tilde{\psi}_{0}} &= e^{\frac{1}{2} \mathbf{\hat{\tau}}^2} \ket{\tilde{\psi}_{0}} \\
    &= \int \frac{d\mathbf{y}}{(2\pi)^{N_g/2}} e^{-\frac{\mathbf{y}^2}{2}} e^{\mathbf{y} \cdot \mathbf{\hat{\tau}}} \ket{\tilde{\psi}_0} \label{eq:hscc}
\end{split}
\end{equation}
Where $\mathbf{y}$ are the auxiliary fields (similar to $\mathbf{x}$ in imaginary-time propagation), and one can directly use (\ref{eq:hscc}) to sample the CCSD wavefunction and evaluate the local energy. However, this leads to a large noise and one can reduce it significantly by using the following correlated sampling

\begin{equation} \label{eq:gaus}
\begin{split}
    &\ket{\Psi_\text{CCSD}} \\
    &= \left[(1 + \hat{T}_2) + (e^{\hat{T}_2} - 1-\hat{T}_2)\right] \ket{\tilde{\psi}_{0}} \\ 
    &= (1 + \hat{T}_2) \ket{\tilde{\psi}_0} \\
    & \quad + \int \frac{d\mathbf{y}}{(2\pi)^{N_g/2}} e^{-\frac{\mathbf{y}^2}{2}} (e^{\mathbf{y} \cdot \mathbf{\hat{\tau}}} - 1 - \mathbf{y}\cdot\mathbf{\hat{\tau}} - \frac{1}{2}(\mathbf{y}\cdot\mathbf{\hat{\tau}})^2 )\ket{\tilde{\psi}_0} \\
    &= (1 + \hat{T}_2) \ket{\tilde{\psi}_0} + \sum_{\mathbf{y}} \ket{\tilde{\psi}_{cr}(\mathbf{y})}
\end{split}
\end{equation}
where we have defined 
\begin{align}
     \ket{\tilde{\psi}_{cr}(\mathbf{y})} = (e^{\mathbf{y} \cdot \mathbf{\hat{\tau}}} - 1 - \mathbf{y}\cdot\mathbf{\hat{\tau}}- \frac{1}{2}(\mathbf{y}\cdot\mathbf{\hat{\tau}})^2 )\ket{\tilde{\psi}_0} 
\end{align}
which only contains contributions from quadruple and higher order excitations that have a small amplitude as long as $T_2$ is less than 1. 
This result follows from the fact that 
 \begin{align}
\int \frac{d\mathbf{y}}{(2\pi)^{N_g/2}} e^{-\frac{\mathbf{y}^2}{2}} ( 1 +\mathbf{y}\cdot\mathbf{\hat{\tau}}+\frac{1}{2}(\mathbf{y}\cdot\mathbf{\hat{\tau}})^2 ) = 1+\hat{T}_2  
 \end{align}

Note that although $\mathbf{y}\cdot\mathbf{\hat{\tau}}$ is zero when integrated against a Gaussian distribution, it cancels the first-order fluctuation of $e^{\mathbf{y}\cdot\mathbf{\hat{\tau}}}$ when the integral in Eq.~\ref{eq:gaus} is sampled using the Monte Carlo technique. Thus, the AFQMC energy with the stochastically sampled CCSD trial can be written as
\begin{equation}
\begin{split}
    &\frac{\braket{\Psi_\text{CCSD}|\hat{H}|\Psi_\text{AF}}}{\braket{\Psi_\text{CCSD}|\Psi_\text{AF}}} \\
    &= \frac{\braket{(1+\hat{T}_2)\tilde{\psi}_0|\hat{H}|\Psi_\text{AF}} + \sum_{\mathbf{y}}\braket{\tilde{\psi}_{cr}(\mathbf{y})|\hat{H}|\Psi_\text{AF}}}{\braket{(1+\hat{T}_2)\tilde{\psi}_0|\Psi_\text{AF}}+ \sum_{\mathbf{y}}\braket{\tilde{\psi}_{cr}(\mathbf{y})|\Psi_\text{AF}}} \\
\end{split}
\end{equation}
The first term and the contribution due to $1  +(\mathbf{y}\cdot\mathbf{\tau})^2$ in $\tilde{\psi}_{cr}(\mathbf{y})$ can be evaluated deterministically using the fast local energy evaluation algorithm of AFQMC/CISD. 

In this formulation, the same as in the other two, we use Hartree-Fock as the guiding wavefunction and the resulting AFQMC energy becomes
\begin{equation}
\begin{split}
    &\frac{\braket{\Psi_\text{CCSD}|\hat{H}|\Psi_\text{AF}}}{\braket{\Psi_\text{CCSD}|\Psi_\text{AF}}} \\
    &= \frac{ \sum_{i\mathbf{y}} \frac{w_i}{\langle\psi_0|\phi_i\rangle} \left[ \braket{(1+\hat{T}_2)\tilde{\psi}_0|\hat{H}|\phi_i} + \braket{\tilde{\psi}_{cr}(\mathbf{y})|\hat{H}|\phi_i} \right] }{ \sum_{i\mathbf{y}} \frac{w_i}{\langle\psi_0|\phi_i\rangle} \left[ \braket{(1+\hat{T}_2)\tilde{\psi}_0|\phi_i} + \braket{\tilde{\psi}_{cr}(\mathbf{y})|\phi_i} \right] }. \label{eq:stoccsd}
\end{split}
\end{equation}
The expression above can be thought of as performing a random walk while sampling $\{x\}$ and $\{y\}$. In practice, for a given time step, we sample $\mathbf{y}$ multiple times ($N_y$) for each walker.

A recent publication by Xiao et al.~\cite{xiao2025sto} follows a similar strategy as ours, but employs a stochastic representation of the exponential wavefunction ansatz without correlated sampling, and uses the stochastically sampled wavefunction as the guiding wavefunction as well. The correlated sampling strategy can be readily transferred to their algorithm and can be used with other trial states, including Slater-Jastrow, CCSD, and VAFQMC\cite{sorella2023systematically,levy2024automatic}. However, as we show in section~\ref{sec:disc}, this method is unlikely to be practical in its current form because we expect the variance in energy to increase exponentially with the size of the system. In that section, we also suggest ways of making it into a practical method using similarity transformation, which we have not pursued in the current publication. For our current paper, this method serves as a reference (for small systems) which helps us gauge the error incurred due to the use of the perturbative expression to evaluate local energy.

In this work, we refer to this flavor as AFQMC/stoCCSD (or STO for short).

\subsection{Discussion}\label{sec:disc}

We begin by discussing the scaling of each method. Benefiting from the fact that the QMC samplings in both PT/PT2 are guided by the HF wavefunction, the cost of the propagation, i.e., calculating the overlap and the force bias, scales as $O(N^3)$ - same as AFQMC/HF. The cost of evaluating the energy for the two perturbative methods scales the same as AFQMC/CISD, $O(N^5)$, due to the presence of $\frac{\braket{\psi_0|\hat{T}^\dagger \hat{H}|\Psi_\text{AF}}}{\braket{\psi_0|\Psi_\text{AF}}}$ in PT and the similar term $\frac{\braket{\tilde{\psi}_0|\hat{T}^\dagger \hat{H}|\Psi_\text{AF}}}{\braket{\tilde{\psi}_0|\Psi_\text{AF}}}$ in PT2. The cost of local energy in AFQMC/stoCCSD is formally the same as the two perturbative methods; however, it also depends on the number of samples, $N_y$, used in Eq.~\ref{eq:stoccsd} per walker. The cost of evaluating the integrant in Eq.~\ref{eq:gaus} scales as $O(N^4)$ since the double excitations, $(\mathbf{y}\cdot\mathbf{\tau})^2 $, are disconnected. Therefore, the final cost of a single energy calculation in STO is $O(N^5) + O(N_y N^4)$. An examination of N$_2$ in the Supporting Information shows that $O(N_y)=1$ is sufficient to capture the CCSD trial without bias for small systems. To see it more clearly, we summarize the cost of single operations in AFQMC/HF, AFQMC/CISD, AFQMC/ptCCSD, AFQMC/pt2CCSD, and AFQMC/stoCCSD in Tab.~\ref{tab:cost}.

It is worth pointing out that although we have derived the local energy expressions starting from CISD wavefunction and performing perturbation theory, it cannot be viewed simply as an approximation to AFQMC/CISD because the force bias and phaseless approximation uses HF trial instead of CISD trial, which makes the cost of propagation cheaper. This does not automatically mean that the AFQMC wavefunction is inferior in the current theories because even if CISD wavefunction is exact (for example in 2-electron systems \cite{li2025fokker}) one is not guaranteed to get an exact AFQMC wavefunction. 

A shortcoming of the PT theory is that it does not follow the zero variance principle when CISD is exact. The energy can still have a bias, although we find that the bias is quite small. An encouraging finding is that the stochastic error in the simulation for all the systems we have studied 
is quite similar to that of an AFQMC/CISD calculation. This implies that in all systems studied in this work (with the exception of H$_2$ where CISD is an exact wavefunction) the cost of the calculation is virtually identical to CISD because the local energy cost is the same and the number of local energy evaluations needed are roughly the same as well. 

Finally, even though the expressions here are obtained using perturbation theory, we do not have any terms that have a difference of orbital energies in the denominator, which means that we do not expect the theory to diverge for metallic systems with vanishing band gaps. This coupled with the size-extensivity of the theory ensures that one can safely use it to obtain energies of metallic systems in the thermodynamic limit. 

Having discussed the scaling of energy with the system size for the various theories, we go on to describe the scaling of the variance (which has implications on the number of samples that one needs to use to get the noise below a user specified threshold) as a function of the system size. 

\begin{table}[]
    \centering
    \begin{tabular}{ccc}
    \hline\hline
Method & Propagation & Energy \\
\hline
AFQMC/HF & $O(N^3)$ & $O(N^4)$  \\
AFQMC/CISD & $O(N^4)$ & $O(N^5)$ \\
AFQMC/ptCCSD & $O(N^3)$ & $O(N^5)$ \\
AFQMC/pt2CCSD & $O(N^3)$ & $O(N^5)$ \\
AFQMC/stoCCSD & $O(N^3)$ & $O(N^5)$ \\
\hline\hline
    \end{tabular}
    \caption{Summary of the cost of a single operation on a walker for all the AFQMC methods used in this work. The first column is the name of each method. The second column is the cost of one propagation step. The last column is the cost of one energy evaluation. The cost due to number of QMC samples required for a desired level of stochastic noise is in the discussion of each method. }
    \label{tab:cost}
\end{table}

\subsubsection{Scaling of Noise of AFQMC/HF with System Size}
The energy estimation by AFQMC/HF and AFQMC/CISD can be written as:
\begin{equation}
    E = \frac{\sum_i w_i E_L(\phi_i)}{\sum_i w_i}
\end{equation}
Where $w_i$ is the weight of walker $i$ and $E_L(\phi_i)$ is the local energy. Due to stochastic reconfiguration, the weights are always $O(1)$. Considering $N$ independent systems, the AFQMC/HF energy can be written as
\begin{equation}
    E = \frac{\sum_i w_i \sum^N_{a=1} E^a_L(\phi^a_i)}{\sum_i w_i} = \sum^N_{a=1} E^a
\end{equation}
The above equation follows from the fact that the local energy of AFQMC/HF is size-extensive and can be written as a sum of $N$ independent systems. Thus, the variance of the total AFQMC/HF energy is the summation of all variances of the energies of each independent system:
\begin{equation}
    \sigma^2_E = \sum_{a=1}^N \sigma^2_{E^a}
\end{equation}
This leads to a $O(N^{1/2})$ dependence of the stochastic noise with system size in AFQMC/HF. To obtain a fixed error, the number of samples needed, therefore, scales as $O(N)$ with system size.

\subsubsection{Scaling of Noise of AFQMC/CISD with System Size}
However, the stochastic noise of AFQMC/CISD acquires a different scaling since the local energy expression couples independent fragments (ignoring single excitations):
\begin{equation}
\begin{split}
    E_L(\phi) =& \frac{\braket{\psi_0|\hat{H}|\phi} + \braket{\psi_0|\hat{C}_2^\dagger \hat{H}|\phi}}{\braket{\psi_0|\phi} + \braket{\psi_0|\hat{C}_2^\dagger|\phi}} \\
    =& \left[ \sum^N_{a=1}  \frac{\braket{\psi^a_0|\hat{H}^a|\phi^a}}{\braket{\psi^a_0|\phi^a}} + \frac{\braket{\psi^a_0|\hat{C}_2^{a\dagger} \hat{H}^a|\phi}}{\braket{\psi^a_0|\phi^a}} \right. \\
    & + \left. \sum^N_{a\neq b} \frac{\braket{\psi^a_0|\hat{C}_2^{a,\dagger}|\phi^a}}{\braket{\psi^a_0|\phi^a}} \frac{\braket{\psi^b_0|\hat{H}^b|\phi^b}}{\braket{\psi^b_0|\phi^b}} \right] \\
    &/ \left[ 1 + \sum^N_{a=1}  \frac{\braket{\psi^a_0|\hat{C}^{a,\dagger}_2|\phi^a}}{\braket{\psi^a_0|\phi^a}} \right]
\end{split}
\end{equation}
where $\hat{C}_2 = \hat{P}_{D} e^{\hat{T}}$ is the double excitation operator obtained by projecting CCSD onto the CISD subspace \cite{mahajan2025afqmc_ci}. Notice that the local energy is not size-extensive and takes the form $X/Y$, where the numerator $X$ contains $N^2$ number of terms and the denominator $Y$ contains $N$ number of terms.

Although no general closed-form expression for the variance of the ratio exists, the practical formula derived from the Taylor series expansion gives
\begin{align}
 \sigma^2_{X/Y} = \frac{\sigma^2_X}{\mu_Y^2} +  \frac{\mu_X^2 \sigma^2_Y}{\mu_Y^4}  - 2\frac{\mu_X \mathrm{cov}(X,Y)}{\mu_Y^3}
\end{align}
where the second term $\frac{\mu_X^2 \sigma^2_Y}{\mu_Y^4}$ scales as $N^2$ with the system size. 
Thus the variance of the local energy scales quadratically with system size.

\subsubsection{Scaling of Noise of AFQMC/ptCCSD with System Size}
For AFQMC/ptCCSD, though the energy is size-extensive, it cannot be split into the summation of $N$ random variables. The two terms in the first-order PT correction, $E^{[1]}$, (Eq.~\ref{eq:ept1}) can be written as:
\begin{equation}
\begin{split}
    & \frac{\braket{\psi_0|\hat{T}^\dagger \hat{H}|\Psi}}{\braket{\psi_0|\Psi}} = \sum^N_{a}\frac{\braket{\psi^N_0|\hat{T}^{a\dagger} \hat{H}^a|\Psi^a}}{\braket{\psi^a_0|\Psi^a}} \\
    & \qquad \qquad  \qquad+ \sum^N_{a\neq b}\frac{\braket{\psi^a_0|\hat{T}^{a\dagger}|\Psi^a}}{\braket{\psi^a_0|\Psi^a}} \frac{\braket{\psi^b_0|\hat{H}^b|\Psi^b}}{\braket{\psi^b_0|\Psi^b}}\\
    &\frac{\braket{\psi_0|\hat{T}^\dagger|\Psi}}{\braket{\psi_0|\Psi}} \frac{\braket{\psi_0|\hat{H}|\Psi}}{\braket{\psi_0|\Psi}} = \sum^N_{a,b} \frac{\braket{\psi^a_0|\hat{T}^{a\dagger}|\Psi^a}}{\braket{\psi^a_0|\Psi^a}} \frac{\braket{\psi^b_0|\hat{H}^b|\Psi^b}}{\braket{\psi^b_0|\Psi^b}} \\ \label{eq:E1}
\end{split}
\end{equation}
Although the difference of
\begin{equation}
    \sum^N_{a\neq b}\frac{\braket{\psi^a_0|\hat{T}^{a\dagger}|\Psi^a}}{\braket{\psi^a_0|\Psi^a}} \frac{\braket{\psi^b_0|\hat{H}^b|\Psi^b}}{\braket{\psi^b_0|\Psi^b}}
\end{equation}
in the two terms in Eq.~\ref{eq:E1} is statistically zero, it still contributes to the variance of $E^{[1]}$. And such variance scales as $O(N^2)$ with system size. We would like to point out that if local correlation methods such as local natural orbitals (LNO) are used, then this quadratic scaling will disappear.

\subsubsection{Scaling of Noise of AFQMC/pt2CCSD with System Size}
The same $N^2$ scaling of variance will be present in AFQMC/pt2CCSD. However, PT2 contains another source of variance hidden in its energy expression - the modified weight 
\begin{equation}
    w' = w \frac{\braket{\psi_T|\phi}}{\braket{\psi_G|\phi}}
\end{equation}
where $\psi_T = \tilde{\psi}_0 = e^{\hat{T}_1}\psi_0$ and $\psi_G = \psi_0$ in PT2 approximation. This overlap ratio introduces a system size-dependent fluctuation in the modified weight. More concretely, since $\frac{\braket{\psi_T|\phi}}{\braket{\psi_G|\phi}}$ is multiplicative
\begin{equation}
    r=\frac{\braket{\psi_T|\phi}}{\braket{\psi_G|\phi}} = \prod^N_a \frac{\braket{\psi^a_T|\phi^a}}{\braket{\psi^a_G|\phi^a}}
\end{equation}
And if we assume that the mean and variance of the overlap ratio of each fragment are $\mu$ and $\sigma^2$ respectively, then the variance of the ratio is
\begin{equation}
    \sigma^2_r = (\mu^2 + \sigma^2)^N - \mu^{2N}\label{eq:singleFragment}
\end{equation}
which increases exponentially with the size of the system. In cases where $\sigma^2 \ll \mu^2$, the variance of the ratio becomes $\sigma^2_2 \approx N \mu^{2N-2} \sigma^2$. When $\mu \approx 1$, as is often the case in chemical systems where $T_1$ amplitudes are small, the practical variance still increases quadratically with system size due to fluctuations in the local energy.

\subsubsection{Scaling of Noise of AFQMC/stoCCSD with System Size}
In AFQMC/stoCCSD, however, the exponential growth of the noise with system size becomes dominant because both $\sigma^2$ and $\mu$ in  (\ref{eq:singleFragment}) are larger, the former due to the fact that the overlap is evaluated stochastically using the Hubbard-Stratonovich decomposition of $e^{\hat{T}_2}$ and the latter because $\hat{T}_2$ amplitudes are typically larger than $\hat{T}_1$. Thus, when performing AFQMC/stoCCSD, we see a dramatic increase in the variance with the size of the system as demonstrated numerically in Fig.~\ref{fig:o2noise}. This makes it unsuitable as a general method; however, for small systems, it still serves as a reference method to gauge the accuracy of the two perturbative approaches.

It is worth pointing out that the variance of AFQMC/stoCCSD can be reduced and made comparable to PT2 by using the similarity transformation as follows
\begin{align}
  \frac{\langle\psi_0|e^{T^\dagger} H |\phi_i\rangle}{\langle\psi_0|\phi_i\rangle} =& \frac{\langle\psi_0|\left(e^{T^\dagger} H e^{-T^\dagger}\right)e^{T^\dagger}|\phi_i\rangle}{\langle\psi_0|\phi_i\rangle} \nonumber\\
  = \frac{\langle\psi_0|H_{\mathrm{sim}}e^{T^\dagger}|\phi_i\rangle}{\langle\psi_0|\phi_i\rangle} \nonumber\\
\end{align}
where $H_{\mathrm{sim}}$ is the similarity-transformed Hamiltonian. Because $T$ only contains excitation operators and $H$ only contains two-body terms, the similarity-transformed Hamiltonian contains at most 6-body terms. This will lead to an algorithm with a dramatic increase in the cost of local energy evaluation but with the benefit that variance will be smaller. 

\section{Results}
\label{sec:numerical}

In this section, we present numerical benchmarks demonstrating the size-extensivity and evaluating the accuracy of our proposed methods: AFQMC/ptCCSD (PT) and AFQMC/pt2CCSD (PT2). For some of the systems, we compare PT/PT2 against AFQMC/stoCCSD (STO) to examine the error introduced by the perturbative approximation. The results are presented in the order:
\begin{enumerate}
    \item \textit{Size-Extensivity in Non-Interacting Systems}: We verify the size-extensivity of PT/PT2 using 1 - 50 H$_2$ and 1 - 5 O$_2$ monomers. These examples also provide evidence that the stochastic noise of PT/PT2 is comparable to AFQMC/CISD, and often much lower than the noise of AFQMC/HF, even for large systems (although in principle eventually the variance of former will become larger than the latter).
    \item \textit{Interacting Systems at the Thermodynamic Limit}: We show that our new size-extensivity approach also provides better results than AFQMC/CISD for interacting systems at the thermodynamic limit (TDL) - 1D Hydrogen atom chain and uniform electron gas (UEG) - without suffering from the infrared divergence.
    \item \textit{Main Group Molecules}: We examine the accuracy of the PT and PT2 on small to medium main group molecules. This includes the ground-state and atomization energy of the systems in the HEAT and W4-MR datasets, the dissociation curve of N$_2$, and the ground-state energy of benzene. 
    \item \textit{3d-Transition Metal Complexes}: We apply PT and PT2 to evaluate the atomization, isomerization, and spin-splitting energy of some 3d-transition metal complexes, demonstrating our methods can be used to study strongly correlated transition metal systems with significant multi-reference characters.
\end{enumerate}
In systems where the Coupled-Cluster singles amplitudes are absent (e.g., H$_2$ and the UEG), PT and PT2 are identical. For the consistency of notation, their results are labeled as AFQMC/pt2CCSD in this work. All plots are color-coded as: CCSD(T) - blue, AFQMC/HF - orange, AFQMC/CISD - green, AFQMC/ptCCSD - red, AFQMC/pt2CCSD - purple, AFQMC/stoCCSD - pink, and black for the reference. Comprehensive numerical data and supporting evidence are provided in the Supplementary Information (SI).

\subsubsection{Non-Interacting Monomers}

\begin{figure}[h]
    \centering
    \includegraphics[width=0.5\textwidth]{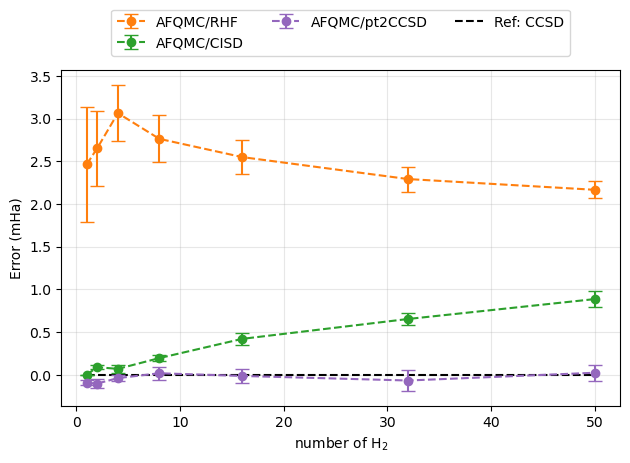}
    \caption{Energy per molecule of non-interacting Hydrogen molecules in STO-6G basis, with bondlength = 1.05835 \AA. The x-axis is the number of monomers. The y-axis is the energy difference from the CCSD reference in mHa. Calculations are in spin-restricted orbitals, and all QMC samplings use 300 walkers and 500 samples. }
    \label{fig:h2}
\end{figure}

\begin{figure}[h]
    \centering
    \includegraphics[width=0.5\textwidth]{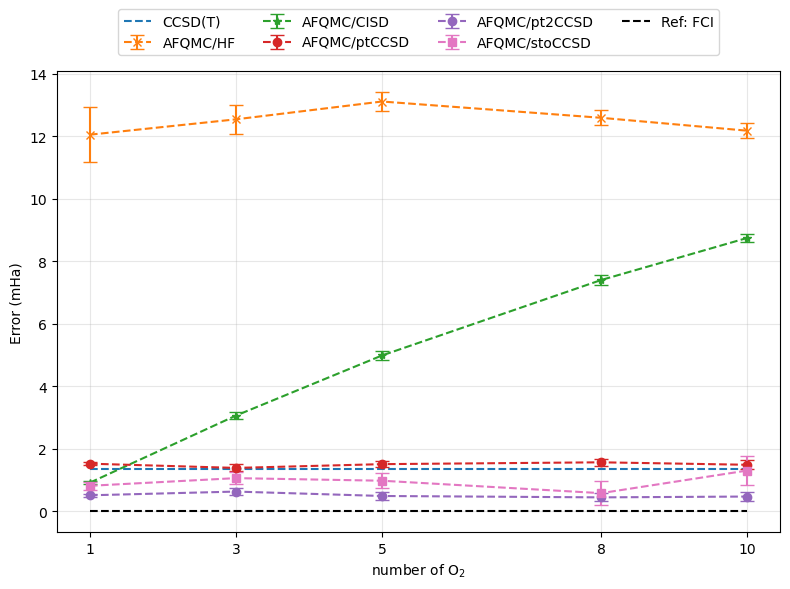}
    \caption{Energy per molecule of non-interacting Oxygen molecules in STO-6G basis, with bondlength = 1.20577 \AA. The x-axis is the number of monomers. The y-axis is the energy difference from the FCI reference. Calculations are in spin-unrestricted orbitals, and all QMC samplings use 300 walkers and 1000 samples.}
    \label{fig:o2}
\end{figure}

This work aims to resolve the lack of size-extensivity in the AFQMC/CISD energy while preserving its high accuracy and low stochastic noise. To demonstrate that PT and PT2 are strictly size-extensive, we present results for two non-interacting systems in the minimum basis (STO-6G): H$_2$ (Fig.~\ref{fig:h2}) and O$_2$ (Fig.~\ref{fig:o2}) molecules. The non-interacting Hydrogen consists of 1, 2, 4, 8, 16, 32, and 50 monomers, with a bond length of 1.05835 \AA. The non-interacting Oxygen consists of 1, 3, 5, 8, and 10 monomers with a bond length of 1.20577 \AA. Simulations for all H$_2$ systems utilized 300 walkers and 500 samples, while all the O$_2$ systems employed 300 walkers and 1000 samples. For O$_2$, we also present the results from AFQMC/stoCCSD with 5 samples of the stochastic trial per walker ($n_s=5$), while the number of walkers and the number of QMC samples are the same as other methods. The calculations for the hydrogen systems are in spin-restricted orbitals, and the calculations of oxygen systems are in spin-unrestricted orbitals.

\begin{figure}[h]
    \centering
    \includegraphics[width=0.5\textwidth]{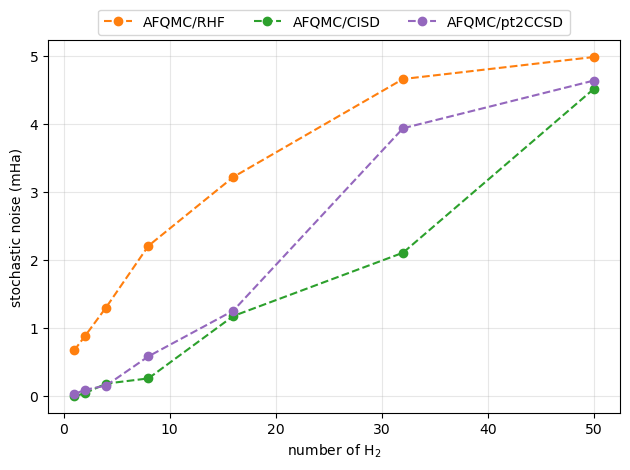}
    \caption{Stochastic noise of non-interacting Hydrogen molecules in STO-6G basis, with bondlength = 1.05835 \AA. The x-axis is the number of monomers. The y-axis is the stochastic noise of each method in mHa. Calculations are in spin-restricted orbitals, and all QMC samplings use 300 walkers and 500 samples.}
    \label{fig:h2noise}
\end{figure}

Coupled-Cluster with singles and doubles (CCSD) provides the exact solution for non-interacting hydrogen molecules, as simultaneous excitations across different molecules are inherently disconnected. AFQMC/CISD is exact for one H$_2$ molecule, due to the zero-variance principle. However, as the system size increases, higher-order excitations become necessary to accurately describe the ground-state wavefunction. The inability of the CISD trial to capture these excitations results in the lack of size-extensivity of AFQMC/CISD \cite{mahajan2025afqmc_ci}. This is clearly demonstrated in Fig.~\ref{fig:h2}, where the AFQMC/CISD energy per molecule deviates from the CCSD reference as the number of monomers increases, reaching an error of approximately 1 mHa per molecule for a 50-monomer system. 

In contrast, the energy per molecule for AFQMC/pt2CCSD remains constant (within stochastic error) as the system size increases, confirming the size-extensivity of the perturbative method 
(Note that there are no single excitations, and therefore PT and PT2 are equivalent). Although the perturbative treatment introduces an approximation that breaks the zero-variance principle when the trial is exact, the error of AFQMC/pt2CCSD from the true ground state is negligible (less than 0.1 mHa per H$_2$).

We also compare the stochastic noise of each method in Fig.~\ref{fig:h2noise}. The noise of AFQMC/HF grows as $O(N^{\frac{1}{2}})$ whereas the noise of the other two methods increases linearly with system size, consistent with the theoretical analysis in the Discussion section \ref{sec:disc}. For the smaller system (1 - 16 H$_2$), the stochastic noise associated with PT2 is much lower than that of AFQMC/HF. Promisingly, AFQMC/pt2CCSD does not exhibit a noticeably higher variance than AFQMC/CISD.

\begin{figure}[h]
    \centering
    \includegraphics[width=0.5\textwidth]{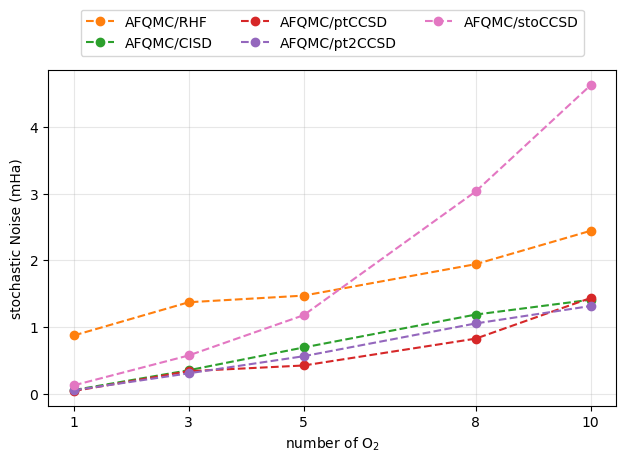}
    \caption{Total stochastic noise of $N$ non-interacting O$_2$ molecules. Basis set = STO-6G basis, bondlength = 1.20577 \AA. The x-axis is the number of monomers. The y-axis is the stochastic noise (mHa). Calculations are in spin-unrestricted orbitals, and all QMC samplings use 300 walkers and 1000 samples. 5 samples of the stochastic trial are used per walker in STO. }
    \label{fig:o2noise}
\end{figure}

The absence of single excitations in the previous example makes that system inadequate for comparing PT and PT2. Therefore, we switch to non-interacting Oxygen molecules in which the Coupled-Cluster singles amplitudes ($T_1$) are more significant due to the spin-triplet open-shell electronic ground state. Fig.~\ref{fig:o2} shows that the energies of both PT and PT2 remain constant as more O$_2$ molecules are included. Conversely, the lack of size-extensivity in AFQMC/CISD results in an error of approximately 5 mHa per molecule in a system of just 5 monomers. Notably, PT2 yields results that are closer to both the FCI and the STO references, suggesting our exact treatment of $T_1$ is necessary. However, in this case, treating the CCSD doubles amplitudes exactly offers little improvement over the PT2 approximation. 

Fig.~\ref{fig:o2noise} demonstrates that the stochastic noise of AFQMC/CISD and the two perturbative approaches are comparable to each other and significantly lower than that of AFQMC/HF. The noise of the STO method, however, increases dramatically with system size, rendering it impractical for larger systems.

\subsubsection{Interacting Systems and The Thermodynamic Limit}
Having demonstrated that PT and PT2 energies are extensive in the last section, as both methods become significantly more accurate than AFQMC/CISD when many monomers are considered simultaneously, it is pertinent to examine whether this accuracy can be extended into interacting systems at the thermodynamic limit (TDL). For a theory/method to be reliable for real materials and condensed-matter systems, it must not only be size-extensive, but it is also desirable to be robust against the infrared divergence for systems with low-energy excitations. To address these questions, we study two toy models: the 1D hydrogen atomic chain in a minimum basis (STO-6G) and the Uniform Electron Gas (UEG).

\paragraph{One-dimensional Hydrogen Atomic Chain} in the minimum basis (STO-6G) serves as a real-world counterpart of the 1D half-filled Hubbard model that allows long-range hopping and off-site electron-electron interactions. In this study, we utilize open boundary conditions with a fixed interatomic separation of 2 Bohr. The size of the chains considered in this work includes N = 4, 8, 16, 32, 64, and 100 atoms. Density Matrix Renormalization Group (DMRG) is used as the reference since it provides near-exact solutions for 1D systems \cite{motta2017hchain, motta2020hchain}. All QMC samplings are performed with 300 walkers and 1000 samples. All calculations are performed in the spin-restricted orbitals.

\begin{figure}[h]
    \centering
    \includegraphics[width=0.5\textwidth]{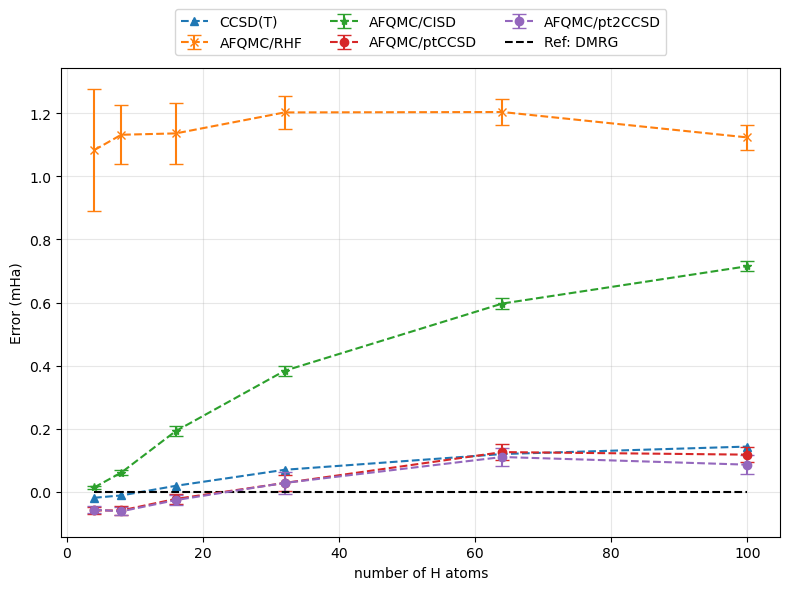}
    \caption{The energy per atom versus the number of atoms by each method (STO-6G basis). The y-axis is the error (mHa) with respect to the DMRG reference. The x-axis is the number of H atoms. Hydrogen atoms are separated by 2 Bohr. All QMC samplings use 300 walkers and 1000 samples. All calculations use RHF-based trial states.}
    \label{fig:hchain}
\end{figure}

\begin{figure}[h]
    \centering
    \includegraphics[width=0.5\textwidth]{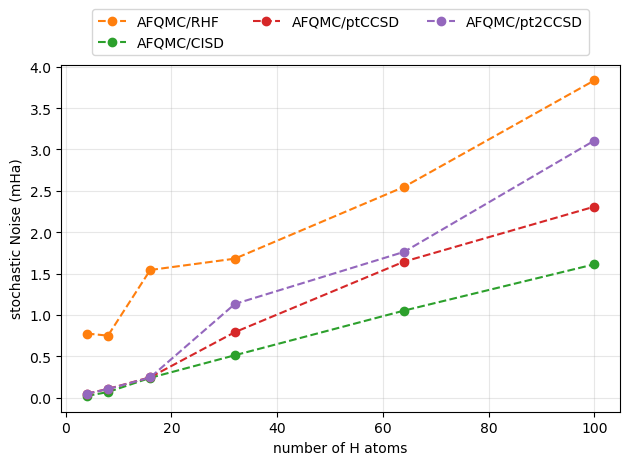}
    \caption{The stochastic noise of the total system versus the number of atoms by each method (STO-6G basis). The y-axis is the stochastic noise (mHa). The x-axis is the number of H atoms. Hydrogen atoms are separated by 2 Bohr. All QMC samplings use 300 walkers and 1000 samples. All calculations are in spin-restricted orbitals.}
    \label{fig:hchain_noise}
\end{figure}

Fig.~\ref{fig:hchain} illustrates the energy per hydrogen atom as a function of the number of atoms in each chain. Compared to the non-interacting monomers in Fig.~\ref{fig:h2}, the energy per atom here is not constant due to the additional interaction. Nevertheless, the advantages of size-extensivity remain clear: the AFQMC/CISD energy deviates increasingly from the DMRG reference as the chain grows - approaching 0.7 mHa error per atom - whereas PT and PT2 yield results comparable to CCSD(T), maintaining an accuracy within approximately 0.1 mHa per atom relative to DMRG. We also note that if the spin-unrestricted HF state is employed as the trial, AFQMC/UHF already yields energy comparable to DMRG, which covers the lack of size-extensivity of AFQMC/UCISD in this example.

The total stochastic noise of each AFQMC method is provided in Fig.~\ref{fig:hchain_noise}. It is found that all methods exhibit a linear growth of the noise versus system size. For smaller chains, the stochastic noise levels of PT, PT2, and AFQMC/CISD are comparable and significantly lower than those of AFQMC/HF. As the system size increases, the noise in the PT and PT2 exceeds that of AFQMC/CISD, though it remains notably lower than the noise associated with AFQMC/HF.

Although CCSD(T) yields accuracy comparable to PT/PT2 in this numerical simulation of hydrogen chains, the perturbative triples correction is known to fail catastrophically at the thermodynamic limit (TDL) in systems with low-energy excitations, where the band gap closes \cite{masios2023ccsdct}. To determine if our proposed approaches are robust against the infrared divergence, we further evaluate their performance using the uniform electron gas.

\paragraph{Uniform Electron Gas} (UEG), also known as homogeneous/degenerate electron gas or jellium, is a fundamental model for many-electron systems and a cornerstone for density-functional theory (DFT) \cite{giuliani2008quantumliquid, loos2016ueg, ceperley1980ueg, kohn1965self, jones2015density}. It consists of interacting electrons moving in a uniform and positively charged background that ensures overall charge neutrality. Under the periodic boundary conditions, the UEG model Hamiltonian is:
\begin{equation}
\begin{split}
    H = & E_M + \sum_{\mathbf{k}}\frac{\mathbf{k}^2}{2} \hat{a}_{\mathbf{k}}^\dagger \hat{a}_{\mathbf{k}} \\
    &+ \frac{1}{2\Omega} \sum_{\mathbf{q}\neq0} \sum_{\mathbf{k_1}, \mathbf{k_2}} \frac{4\pi}{\mathbf{q}^2} \hat{a}_{\mathbf{k_1}+\mathbf{q}}^\dagger \hat{a}_{\mathbf{k_2}-\mathbf{q}}^\dagger \hat{a}_{\mathbf{k_2}} \hat{a}_{\mathbf{k_1}}
\end{split}
\end{equation}
Where 
$E_M = -2.837297 (\frac{3}{4\pi})^{\frac{1}{3}} N^{\frac{2}{3}} r_s^{-1} $ 
is the Madelung energy~\cite{schoof2015ueg} which accounts for the interaction of each charge with its periodic images. $N$ is the number of electrons per cell, $\Omega$ is the volume of a cell, and $r_s$ is the Wigner-Seitz radius that characterizes inter-electron distance such that $\Omega = \frac{4 \pi}{3}r_s^3N$. 

For the high-density region ($r_s < 2$), the electron gas is weakly correlated; the mean-field approximation performs well and approaches the exact solution as $r_s \rightarrow 0$. In this region, AFQMC/HF provides near-exact benchmarking results\cite{lee2019ueg_afqmc, lee2021ueg_afqmc}. However, as the electron gas becomes less dense at the larger $r_s$ region, electron-electron correlation becomes more significant, and using the HF wavefunction as the trial becomes inadequate \cite{lee2019ueg_afqmc}.

Despite the conceptual simplicity of the UEG, its long-wavelength excitations pose significant challenges for many-body perturbation theories and related computational methods. It is well known that the second-order Møller-Plesset (MP2) diverges as the band gap narrows \cite{shepherd2013ueg_cc}. While Coupled-Cluster theory remains applicable in the thermodynamic limit, perturbative corrections — such as those in CCSD(T) — exhibit problematic behavior \cite{shepherd2013ueg_cc, neufeld2017ueg_cc, neufeld2023ueg_cc, shepherd2014ueg_cc}. Recent efforts have been made to reformulate the perturbative Triple correction to alleviate the infrared divergence, which gives rise to a method called CCSD(cT)\cite{masios2023ccsdct, schafer2024ground}. Consequently, we employ the UEG to rigorously test whether our newly proposed methods can successfully model systems dominated by low-energy excitations without suffering from divergent artifacts.

First, we test the accuracy of AFQMC/pt2CCSD on a 14-electron UEG system in both the weakly correlated ($r_s = 2$) and the more strongly correlated ($r_s = 5$) regions. Tab.~\ref{tab:ueg_basis} summarizes the correlation energies extrapolated to the complete basis set (CBS) limit of these two cases. The CBS extrapolation is performed via an inverse linear fit with M = 179, 257, and 389 basis functions (plane-waves). Both systems are prepared in a spin-restricted HF state (RHF).

\begin{table}[h]
    \centering
    \begin{tabular}{cccc}
    \hline\hline
Method          & $r_s = 2$  &  $r_s = 5$ \\
\hline
CCSD$^a$        & -0.4094(1) & -0.2531(3) \\
CCSD(cT)$^b$    & -0.4382    & -0.2954  \\
CCSDT$^a$       & -0.4354(4) & -0.2970(4) \\
FCIQMC$^c$      & -0.4447(4) & -0.306(1)  \\
AFQMC/HF        & -0.443(2)  &   N/A$^d$  \\
AFQMC/CISD      & -0.444(1)  &   N/A$^d$  \\
AFQMC/pt2CCSD   & -0.444(1)  & -0.289(1)  \\ 
\hline\hline
    \end{tabular}
    \caption{CBS ground-state correlation energy (Ha) of the spin-restricted 14-electron UEG at $r_s$ = 2 and $r_s$ = 5. (a) CCSD and CCSDT results are from the work of Neufeld and Thom \cite{neufeld2017ueg_cc}. (b) CCSD(cT) results are from the work of Masios et al.\cite{masios2023ccsdct} 
    (c) FCIQMC data are from the work of Shepherd et al.~\cite{shepherd2012ueg_ifciqmc}. (d) Energy behaves non-monotonically with an increasing number of basis functions, as shown in Fig.~\ref{fig:rs5_nk}.}
    \label{tab:ueg_basis}
\end{table}

\begin{figure}[h]
    \centering
    \includegraphics[width=0.5\textwidth]{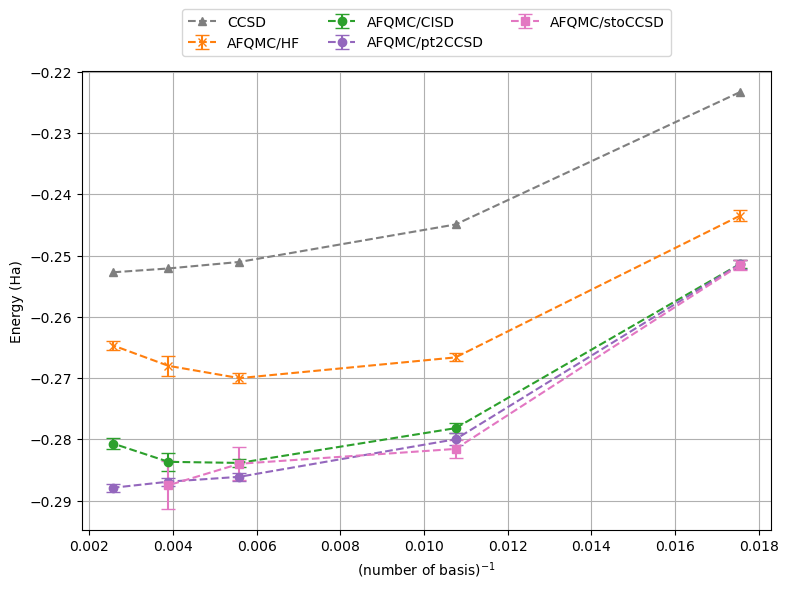}
    \caption{Ground-state correlation energy (Ha) of the 14-electron UEG ($r_s=5$) versus the inverse of the number of basis functions (1/M). The corresponding number of basis functions plotted are M = 57, 93, 179, 257, and 389.}
    \label{fig:rs5_nk}
\end{figure}

For the weakly correlated UEG ($r_s=2$), all three AFQMC methods yield highly accurate results that are comparable to the FCIQMC energy. However, in the more strongly correlated case ($r_s=5$), the energy of AFQMC/HF behaves non-monotonically with respect to basis set size, which was also previously observed by Lee et al.\cite{lee2019ueg_afqmc}. Although Lee et al. suggested that an improved trial wavefunction might alleviate this issue, we find the AFQMC/CISD energy also exhibits non-monotonic increases once the number of basis functions exceeds M = 257, despite capturing more correlation than AFQMC/HF. Fortunately, AFQMC/pt2CCSD does not suffer from this non-monotonic behavior, as illustrated in Fig.~\ref{fig:rs5_nk}; however, it still differs from the FCIQMC reference by approximately 15 mHa.

To examine whether this relatively large deviation of PT2 from the FCIQMC could be an error from the perturbative treatment of the CCSD trial, we compare the PT2 results with those obtained by STO. As depicted by the pink and purple data points in Fig.~\ref{fig:rs5_nk}, these two methods produce nearly identical energies for M = 57, 93, 179, and 257. This agreement strongly suggests that the error introduced by the perturbative approximation in PT2 is negligible.

Next, we examine the behavior of AFQMC/pt2CCSD energy as the band gap of the $r_s = 5$ UEG model closes - the HOMO-LUMO gap decreases with increasing unit cell size. MP2 and CCSD(T) are known to diverge in this condition \cite{shepherd2013ueg_cc, masios2023ccsdct}. For these calculations, the momentum cutoff is maintained at a constant ratio relative to the Fermi momentum, $k_{cut} = \sqrt{2}k_{f}$.

\begin{figure}[h]
    \centering
    \includegraphics[width=0.5\textwidth]{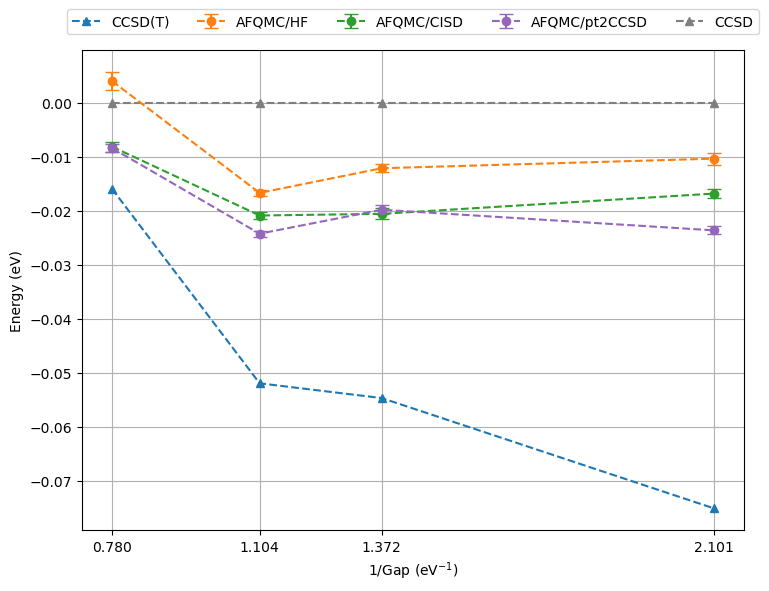}
    \caption{Energy difference (eV) per electron relative to CCSD for the $r_s = 5$ UEG, plotted against the inverse of the HOMO-LUMO band gap (eV$^{-1}$). The momentum cutoff is set to $k_{cut} = \sqrt{2}k_f$. }
    \label{fig:ueg_gap}
\end{figure}

Fig.~\ref{fig:ueg_gap} illustrates the energy difference between each method and CCSD as a function of the inverse of the HOMO - LUMO band gap. The infrared divergence of the perturbative triples (T) correction is clearly demonstrated in this plot as the blue dots diverge rapidly from the CCSD energy. In contrast, all three AFQMC methods successfully capture additional correlation beyond CCSD without exhibiting this unphysical divergence. While the numerical difference between AFQMC/CISD and AFQMC/pt2CCSD is less pronounced here than in our earlier simulations, we anticipate that the rigorous size-extensivity of AFQMC/pt2CCSD will render it increasingly advantageous as system sizes grow toward the thermodynamic limit.

\subsubsection{Main Group Molecular Systems}
Having demonstrated the advantages of our new size-extensive methods on toy models, we now turn to their application in realistic systems. In this section, we comprehensively benchmark the accuracy of both methods on small-to-medium-sized main group chemical systems. We do not expect PT/PT2 to outperform AFQMC/CISD in this regime, as size-extensivity is not a dominant factor for systems at this scale. Moreover, because the phaseless approximation employs HF rather than CISD wavefunctions in this work, the QMC walkers may be less accurate than those of AFQMC/CISD, especially for strongly correlated systems.
Therefore, results comparable to, or marginally inferior to, those of AFQMC/CISD are acceptable for the size-extensive approach. The primary objective of this section is to quantify the deviation between PT/PT2 and AFQMC/CISD trial for small-to-medium systems. 
The examined systems include the HEAT (High-Accuracy Extrapolated Ab Initio Thermochemistry) dataset\cite{tajti2004heat, bomble2006high, harding2008high, thorpe2019high}, the W4-MR dataset\cite{karton2017w4}, the ground state energy of benzene, and the nitrogen bond dissociation curve.

\paragraph{HEAT Dataset} comprises 26 small molecules, and was developed by Stanton and colleagues to study thermochemistry such as molecule formation and atomization with highly accurate Coupled-Cluster methods\cite{tajti2004heat, bomble2006high, harding2008high, thorpe2019high}. Over time, it has become a standard benchmarking tool for testing first-principles theoretical chemistry tools. In this work, we calculate the ground-state and atomization energies of all molecules in the HEAT dataset with the cc-pVDZ basis set\cite{dunning1989gaussian}. The CCSD(T), AFQMC/HF, and AFQMC/CISD results are taken from our group's recent publication \cite{mahajan2025afqmc_ci}. The Coupled-Cluster with Singles Doubles Triples Quadruples and Pentacles (CCSDTQP) energies serving as the references are taken from the work of Bomble et al.\cite{bomble2005ccsdtpq}, except for CO$_2$, whose CCSDTQP energy is not reported therein and is instead taken from Ref.~\citenum{mahajan2025afqmc_ci}. The root-mean-square deviations (RMSD) from the reference, across the whole dataset, are reported in Tab.\ref{tab:heat} and Tab.\ref{tab:heat_atom} for each method. The energy of each molecule is provided in the Supporting Information.

\begin{table}[h]
    \centering
    \begin{tabular}{cccc}
    \hline\hline
Method &  & RMSD (mHa) \\
\hline
CCSD(T)$^a$          & & 1.79 \\
AFQMC/HF$^a$         & & 2.8(2) \\
AFQMC/CISD$^a$       & & 0.75(7) \\
AFQMC/ptCCSD         & & 1.4(1) \\ 
AFQMC/pt2CCSD        & & 1.1(1) \\ 
AFQMC/stoCCSD        & & 1.0(2) \\ 
\hline\hline
    \end{tabular}
    \caption{Ground state energy RMSD (mHa) from CCSDTQP \cite{bomble2005ccsdtpq, mahajan2025afqmc_ci} of molecules in the HEAT dataset (cc-pVDZ basis). (a) The results of CCSD(T), AFQMC/HF, AFQMC/CISD, and the CCSDTQP energy of CO$_2$ are from Ref.~\citenum{mahajan2025afqmc_ci}. }
    \label{tab:heat}
\end{table}

As shown in Tab.~\ref{tab:heat}, AFQMC/CISD achieves the lowest RMSD from the reference across all molecules in this dataset. AFQMC/pt2CCSD is slightly less accurate than AFQMC/CISD, yet it improves upon AFQMC/ptCCSD and surpasses CCSD(T) in ground-state energy accuracy. Notably, AFQMC/pt2CCSD and AFQMC/stoCCSD yield statistically identical results.  Inspection of the per-molecule data in the Supporting Information reveals that the improvement of PT2 over PT is largely attributable to the outlying energy of CN, and that PT2 and AFQMC/stoCCSD are in close agreement throughout the dataset. This again suggests that incorporating $e^{\hat{T}_1}$ exactly is useful, whereas a non-perturbative treatment of the Coupled-Cluster Doubles offers little additional benefit.

We also report the RMSD of the atomization energy (cc-pVDZ basis) in Tab.~\ref{tab:heat_atom}. The atomization energy is defined as $\Delta E=\sum_{a\in m}E_a - E_m$, where $E_m$ is the ground state energy of the molecule and $E_a$ is the ground state energy of an atom in the molecule.

\begin{table}[h]
    \centering
    \begin{tabular}{cccc}
    \hline\hline
Method              & & RMSD (mHa) \\
\hline
CCSD(T)$^a$         & & 1.42 \\
AFQMC/CISD$^a$      & & 0.55(7) \\
AFQMC/ptCCSD        & & 1.24(9) \\ 
AFQMC/pt2CCSD       & & 0.9(1) \\
AFQMC/stoCCSD       & & 1.0(2) \\ 
\hline\hline
    \end{tabular}
    \caption{Atomization energy RMSD (mHa) from CCSDTQP \cite{bomble2005ccsdtpq, mahajan2025afqmc_ci} of molecules in the HEAT dataset (cc-pVDZ basis). (a) The results of CCSD(T), AFQMC/CISD, and the CCSDTQP energy of CO$_2$ are from Ref.~\citenum{mahajan2025afqmc_ci}. }
    \label{tab:heat_atom}
\end{table}

Similar to the results of ground-state energy, the newly proposed methods outperform CCSD(T), and PT2 improves upon PT. PT2 and STO still remain in close agreement.  The residual deviation of PT2 from AFQMC/CISD — on the order of 0.3–0.4 mHa — is acceptable, and the total error (0.9 mHa) with respect to CCSDTQP remains within chemical accuracy.

\paragraph{W4-Multi-Reference Dataset} was generated by Karton et al. using the first-principles Weizmann-4 computational thermochemistry protocol \cite{karton2011w4, karton2017w4}. Compared to the HEAT dataset, it consists of larger and more strongly correlated molecules with significant multi-reference characters. These systems are often considered challenging for many high-level theoretical chemistry methods\cite{lyakh2012multireference}. In the recent study that developed AFQMC/CISD algorithm\cite{mahajan2025afqmc_ci}, both AFQMC with HF and CISD trial wavefunctions were found to produce more accurate results than CCSD(T) on this dataset. To assess whether our new size-extensive methods can preserve their accuracy for multi-reference systems, we compute ground-state and atomization energies (cc-pVDZ basis) using AFQMC/ptCCSD and AFQMC/pt2CCSD. As in the HEAT dataset, AFQMC/stoCCSD serves as a reference to PT2. 
The CCSD(T), AFQMC/HF, AFQMC/CISD, and reference energies are taken from Ref.~\citenum{mahajan2025afqmc_ci}. Note that the reference energies are calculated at the CCSDTQ level for most molecules, while AFQMC/HCI is used for ClO$_3$.
\begin{table}[h]
    \centering
    \begin{tabular}{cccc}
    \hline\hline
Method & RMSD (mHa) \\
\hline
CCSD(T)$^a$         & 4.1 \\
AFQMC/HF$^a$        & 3.3(4) \\
AFQMC/CISD$^a$      & 1.9(2) \\
AFQMC/ptCCSD        & 4.1(2) \\ 
AFQMC/pt2CCSD       & 1.7(2) \\
AFQMC/stoCCSD       & 2.3(4) \\ 
\hline\hline
    \end{tabular}
    \caption{Ground-state energy RMSD (mHa) from the reference of molecules in the W4-MR dataset (cc-pVDZ basis). CCSDTQ is used as the reference for all molecules except ClO$_3$, for which AFQMC/HCI is used. (a) The results of CCSD(T), AFQMC/HF, AFQMC/CISD, and the reference energies are from Ref.~\citenum{mahajan2025afqmc_ci}. }
    \label{tab:w4mr}
\end{table}

Tab.~\ref{tab:w4mr} reports the ground-state energy RMSDs for all molecules in the W4-MR dataset, and Tab.~\ref{tab:w4mr_atom} reports the corresponding atomization energy RMSDs. Per-molecule details are provided in the Supporting Information.

\begin{table}[h]
    \centering
    \begin{tabular}{cccc}
    \hline\hline
Method & & RMSD (mHa) \\
\hline
CCSD(T)$^a$         & & 3.37 \\
AFQMC/CISD$^a$      & & 1.6(2) \\
AFQMC/ptCCSD      & & 4.0(2) \\ 
AFQMC/pt2CCSD     & & 1.5(2) \\ 
AFQMC/stoCCSD     & & 2.1(4) \\ 
\hline\hline
    \end{tabular}
    \caption{Atomization energy RMSD (mHa) from the reference of molecules in the W4-MR dataset (cc-pVDZ basis). CCSDTQ is used as the reference for all molecules except ClO$_3$, for which AFQMC/HCI is used. (a) The results of CCSD(T), AFQMC/HF, AFQMC/CISD, and the reference energies are from Ref.~\citenum{mahajan2025afqmc_ci}. }
    \label{tab:w4mr_atom}
\end{table}

For both the ground-state and the atomization energies on the W4-MR dataset, the benefit of treating the Coupled-Cluster $T_1$ amplitudes exactly rather than perturbatively is even more pronounced compared to the HEAT dataset. In both Tab.~\ref{tab:w4mr} and Tab.~\ref{tab:w4mr_atom}, the RMSD decreases by approximately 2.5 mHa going from PT to PT2. More remarkably, PT2 achieves comparable accuracy to AFQMC/CISD, with both methods outperforming CCSD(T). However, the result of STO again demonstrates that treating $e^{\hat{T}_2}$ exactly offers no additional benefit.

\paragraph{Benzene} is one of the most iconic molecular structures in chemistry, characterized by its six-membered aromatic ring. In a blind challenge\cite{eriksen2020benzene} - in which no reference energy was available - large discrepancies were found even among some of the most accurate methods, including CCSDTQ\cite{oliphant1991ccsdtq, kucharski1992coupled, matthews2015non}, SHCI\cite{petruzielo2012semistochastic, holmes2016efficient, sharma2017shci}, FCIQMC\cite{booth2009fermion, cleland2010fciqmc, ghanem2019unbiasing}, DMRG\cite{white1992density,white1999ab,chan2011density,sharma2012spin,olivares2015ab}, etc. Here, we report the ground-state energy of benzene using the same geometry, basis set (cc-pVDZ), and frozen-core (30 electrons and 108 basis) as in the blind challenge. The ground-state correlation energies of CCSD(T), AFQMC/CISD, AFQMC/ptCCSD, and AFQMC/pt2CCSD are listed in Tab.~\ref{tab:benzene}. CCSD(T), AFQMC/CISD, AFQMC/ptCCSD, and AFQMC/pt2CCSD calculations are performed in this work with spin-restricted orbitals.

\begin{table}[h]
    \centering
    \begin{tabular}{cccc}
    \hline\hline
Method & & Energy (mHa) \\
\hline
CCSD(T)         & & -859.5 \\
AFQMC/CISD      & & -862.0(7) \\
AFQMC/ptCCSD    & & -861.8(4) \\
AFQMC/pt2CCSD   & & -861.8(4) \\
CCSDTQ$^a$      & & -862.4 \\            
DMRG$^a$        & & -862.8 \\
\hline\hline
    \end{tabular}
    \caption{Ground-state correlation energy of Benzene in cc-pVDZ basis. Spin-restricted orbitals are used for the HF state. (a) The CCSDTQ and DMRG energies are taken from the blind challenge \cite{eriksen2020benzene}. }
    \label{tab:benzene}
\end{table}

As shown in Tab.~\ref{tab:benzene}, all three QMC methods yield consistent results within 1 mHa from the DMRG and CCSDTQ reference, substantially outperforming the accuracy of CCSD(T).

\begin{figure}[h]
    \centering
    \includegraphics[width=0.5\textwidth]{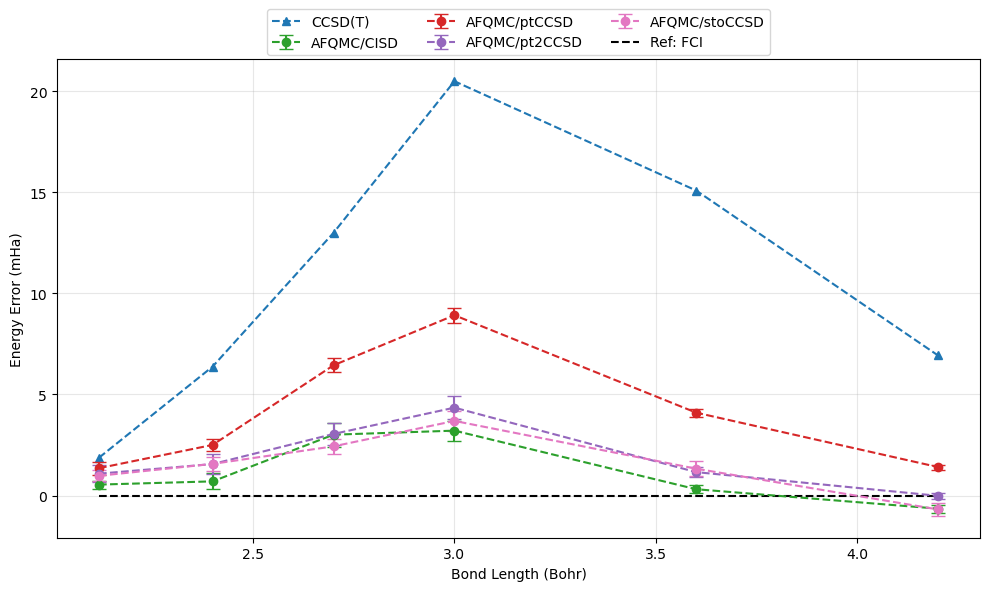}
    \caption{Error in ground-state energy (mHa) with respect to the FCI reference at different N--N bond lengths (cc-pVDZ basis). The FCI reference energy is taken from Chan et al.~\cite{chan2004n2}. CCSD(T), AFQMC/HF, and AFQMC/CISD energies are taken from Ref.~\citenum{mahajan2025afqmc_ci}. }
    \label{fig:n2_diss}
\end{figure}

\paragraph{N$_2$ Dissociation} and bond breaking in general represent a prototypical situation in which multi-reference character and near-degeneracy pose severe difficulties for many electronic structure methods\cite{lyakh2012multireference, bulik2015can}. The exceptionally strong triple bond in the nitrogen molecule makes it the most abundant and stable diatomic molecule in Earth's atmosphere, yet at the same time, a stringent test for theoretical/computational modeling. N$_2$ bond breaking is also of central interest to modern chemistry and the chemical industry. These features have made N-N dissociation a canonical benchmark for highly accurate first-principles methods. In this section, we study the N$_2$ dissociation curve (cc-pVDZ basis) using AFQMC/ptCCSD, AFQMC/pt2CCSD, and AFQMC/stoCCSD. The FCI reference energies are taken from Chan et al.~\cite{chan2004n2}, and the CCSD(T), AFQMC/HF, and AFQMC/CISD energies are taken from Ref.~\cite{mahajan2025afqmc_ci}. All calculations are performed in spin-unrestricted orbitals. The N--N bond lengths considered are 2.118, 2.4, 2.7, 3.0, 3.6, and 4.2~Bohr. Results are shown in Fig.~\ref{fig:n2_diss} and Tab.~\ref{tab:n2_diss}.

As seen in Fig.~\ref{fig:n2_diss}, CCSD(T) significantly underestimates the correlation at the intermediate region ($d \approx 3$~Bohr). This is expected, as the perturbative Coupled-Cluster Triples are known to be unreliable when near-degeneracy occurs \cite{bulik2015can, masios2023ccsdct}. To quantitatively compare the accuracy of each method across the N$_2$ dissociation curve, we report both the RMSD and the non-parallelity error (NPE) in Tab.~\ref{tab:n2_diss}. The NPE is defined as the absolute value of the energy error maximum minus the minimum of a method, $NPE=|E^{error}_{max} - E^{error}_{min}|$, so that a constant shift in the potential surface is not counted.

\begin{table}[h]
    \centering
    \begin{tabular}{ccc}
    \hline\hline
Method              & RMSD (mHa)  & NPE (mHa) \\
\hline
CCSD(T)$^a$         &    12.3     & 18.6 \\
AFQMC/CISD$^a$      &    1.9(4)   & 3.9(5) \\
AFQMC/ptCCSD        &    5.0(3)   & 7.6(5) \\ 
AFQMC/pt2CCSD       &    2.3(4)   & 4.4(6) \\
AFQMC/stoCCSD       &    2.0(4)   & 4.4(6) \\
\hline\hline
    \end{tabular}
    \caption{The RMSD, with respect to the FCI reference, and the NPE at different N--N bond lengths in cc-pVDZ basis. The FCI reference energy is taken from Chan et al.~\cite{chan2004n2}. CCSD(T), AFQMC/HF, and AFQMC/CISD energies are taken from Ref.~\citenum{mahajan2025afqmc_ci}. }
    \label{tab:n2_diss}
\end{table}

As shown in Tab.~\ref{tab:n2_diss}, CCSD(T) performs the worst, with RMSD = 12.3~mHa and NPE = 18.6~mHa, far outside the threshold of chemical accuracy. PT2 substantially improves upon PT and achieves an accuracy comparable to AFQMC/CISD across the entire dissociation curve. Consistent with the results in previous sections, the difference between AFQMC/pt2CCSD and AFQMC/stoCCSD is negligible in both Fig.~\ref{fig:n2_diss} and Tab.~\ref{tab:n2_diss}, indicating that incorporating the full coupled-cluster doubles beyond the PT2 level yields no appreciable improvement.

\subsubsection{3d Transition Metal Complexes}

Beyond the main-group systems, transition metals present another significant challenge for both the chemistry and physics communities. The competition between the kinetic term and electron-electron repulsion in the partially filled $d$ orbitals gives rise to interesting phenomena such as Mott insulator transition\cite{mott1949mott_insul, imada1998metal} and high-T$_c$ superconductors \cite{anderson1987super, muller1987discovery}. In molecular systems, transition metals often play a crucial role serving as the catalytic center of many important chemical reactions \cite{noyori2002asymmetric, liu2018metal, twilton2017merger}. However, the diffusive and strong multi-reference character of the $d$ orbitals makes accurate theoretical and computational description particularly challenging, sustaining an active and ongoing area of research \cite{cramer2009density, szalay2012multiconfiguration}. In this section, we test our new methods on a set of $3d$ transition metal complexes.

\paragraph{Transition Metal Monoxides Molecules} studied in this paper follows the same geometry as the publication by Williams et al.\cite{williams2020tmos} Trial-Needs pseudopotential and corresponding basis set \cite{trail2015correlated, trail2017shape} is used. In Tab.~\ref{tab:tmo}, we report the RMSD of the dissociation energy with respect to the SHCI reference of the 7 monoxides. The dissociation energy is extrapolated to CBS via an inverse cubic fit of TZ and QZ basis. All calculations are performed in spin-unrestricted orbitals. The CCSD(T), AFQMC/HF, AFQMC/CISD, and SHCI reference energies are taken from Ref.~\citenum{mahajan2025afqmc_ci}.

\begin{table}[h]
    \centering
    \begin{tabular}{cc}
    \hline\hline
Method & RMSD (mHa) \\
\hline
CCSD(T)$^a$        &  2.85 \\
AFQMC/HF$^a$       &  10(2) \\
AFQMC/CISD$^a$     &  2(1) \\
AFQMC/ptCCSD       &  6(2) \\ 
AFQMC/pt2CCSD      &  5(2) \\ 
\hline\hline
    \end{tabular}
    \caption{The RMSD with respect to SHCI energies of the CBS extrapolated (TZ - QZ) dissociation energies of the 7 transition metal monoxides. (a) The CCSD(T), AFQMC/HF, AFQMC/CISD, and SHCI reference energies are taken from Ref.~\citenum{mahajan2025afqmc_ci}. }
    \label{tab:tmo}
\end{table}

In this data set, PT/PT2 performs worse than both CCSD(T) and AFQMC/CISD, and the improvement of PT2 over PT is considerably less pronounced than in the preceding examples. This degradation of accuracy may be attributed to the HF guiding wavefunctions employed throughout this work to preserve size-extensivity. Since there typically exist many symmetry-broken HF solutions for transition metal complexes, finding the right solution can be challenging. Also, as mentioned by Mahajan et al.\cite{mahajan2025afqmc_ci}, all methods benefit from error cancellation between the energies of the molecules and the atoms. We also note that the electron correlations of transition metal complexes are often considerably stronger in the form of small molecules than in larger clusters or extended crystal cells. 

\paragraph{The Isomerization of Cu$_2$O$_2^{2+}$} between the f = 0 and f = 1 geometry provides a theoretical model for studying the activation of certain enzymes with implications for understanding the mechanism of C--H bond activation and O$_2$ activation \cite{cramer2006cu2o2, solomon2014copper}. Our group has previously reported a near-exact study of this isomerization energy using free-projection AFQMC (fp-AFQMC)\cite{mahajan2021fp_afqmc}, as well as a comprehensive examination using AFQMC/HCI \cite{malone2022ipie}. A more recent study with AFQMC/CISD \cite{mahajan2025afqmc_ci} demonstrated excellent agreement with both benchmarking references. In Tab.~\ref{tab:cu2o2}, we compare the isomerization energy obtained from AFQMC/ptCCSD and AFQMC/pt2CCSD with these previously reported values.  All calculations are performed with an active space of (32e, 108o), using the same basis set and pseudopotential as in Ref.~\citenum{mahajan2021fp_afqmc}. The CCSD(T) energies are from Ref.~\citenum{cramer2006cu2o2}, the AFQMC/HF and AFQMC/CISD energies are from Ref.~\citenum{mahajan2025afqmc_ci}, and the fp-AFQMC energies are from Ref.~\citenum{mahajan2021fp_afqmc}.

\begin{table*}[t]
    \centering
    \begin{tabular}{cccc}
    \hline\hline
Method          & E(f=0)       & E(f=1)   & $\Delta$E \\
\hline
CCSD(T)$^a$         & -542.0885    & -542.1373    & 30.6 \\
AFQMC/HF$^b$        & -542.0966(7) & -542.152(1)  & 34.8(8) \\
AFQMC/CISD$^b$      & -542.0906(9) & -542.1290(9) & 24.1(8) \\
fp-AFQMC$^c$        & -542.0964(7) & -542.1348(7) & 24.1(6) \\
AFQMC/ptCCSD        & -542.0850(8) & -542.116(1)  & 19.5(8) \\ 
AFQMC/pt2CCSD       & -542.0929(8) & -542.1314(7) & 24.2(7) \\ 
\hline\hline
    \end{tabular}
    \caption{Ground-state energies (Ha) of the two isomers of Cu$_2$O$_2^{2+}$ (columns 2 and 3), and the isomerization energy (kcal/mol) $\Delta$E = E(f=0) - E(f=1) (column 4). (a) The CCSD(T) energies are from Cramer et al. \cite{cramer2006cu2o2}. (b) The AFQMC/HF and AFQMC/CISD energies are from Ref.~\citenum{mahajan2025afqmc_ci}.  (c) fp-AFQMC energies are from Mahajan et al.\cite{mahajan2021fp_afqmc}. }
    \label{tab:cu2o2}
\end{table*}

As summarized in Tab.~\ref{tab:cu2o2}, AFQMC/pt2CCSD is in excellent agreement with both fp-AFQMC and AFQMC/CISD. By contrast, CCSD(T) overestimates the isomerization energy by approximately 6~kcal/mol, and AFQMC/HF overestimates it by approximately 10~kcal/mol. While both AFQMC/pt2CCSD and AFQMC/CISD benefit from error cancellation between the two isomers, the absolute energies of AFQMC/pt2CCSD are in even closer agreement with the fp-AFQMC reference.

\paragraph{The Spin-Splitting of Fe(H$_2$O)$_6^{2+}$} is an example of spin crossover, a phenomenon in which a transition metal complex undergoes geometric deformation upon spin excitation \cite{bousseksou2011spin_cross}. Spin crossover has attracted significant interest for its potential applications in the design of molecular switches \cite{varret2002spin_switch} and for elucidating the activation mechanisms of certain enzymes \cite{shaik2005enzyme}. Here, we present a study on the energy difference between the low-spin (LS) singlet and the high-spin (HS) quintet states of Fe(H$_2$O)$_6^{2+}$ using our new methods (Tab.~\ref{tab:fehydro}). The LS and HS geometries are taken from Ref.~\citenum{floser2020fehydro}. All calculations employ a mixed basis set, with cc-pwCVTZ-DK on the Fe atom and cc-pwCVDZ-DK on the O and H atoms. Scalar relativistic effects are included via the X2C Hamiltonian \cite{saue2011relativistic}. An active space of (62e, 235o) is used for all correlated calculations. The CCSD(T), AFQMC/HF, and AFQMC/CISD energies are taken from Ref.~\citenum{mahajan2025afqmc_ci}. We note that no near-exact theoretical or experimental reference is currently available for this system.

\begin{table}[h]
    \centering
    \begin{tabular}{cc}
    \hline\hline
Method              & $\Delta$E \\
\hline
CCSD(T)$^a$         & -37.7 \\
AFQMC/HF$^a$        & -44.2(8) \\
AFQMC/CISD$^a$      & -36.1(9) \\
AFQMC/ptCCSD        & -36.0(6) \\ 
AFQMC/pt2CCSD       & -38.3(5) \\ 
\hline\hline
    \end{tabular}
    \caption{The Spin-splitting gap (kcal/mol) between the LS and HS states of Fe(H$_2$O)$_6^{2+}$, $\Delta$E = E(LS) - E(HS). (a) The CCSD(T), AFQMC/HF, and AFQMC/CISD are taken from Ref.~\citenum{mahajan2025afqmc_ci}. }
    \label{tab:fehydro}
\end{table}

As shown in Tab.~\ref{tab:fehydro}, AFQMC/ptCCSD is in closest agreement with AFQMC/CISD, while AFQMC/pt2CCSD lies approximately 2~kcal/mol below AFQMC/CISD and 0.6~kcal/mol below CCSD(T). Given the stochastic uncertainties and the absence of a near-exact reference, it is reasonable to conclude that CCSD(T), AFQMC/CISD, AFQMC/ptCCSD, and AFQMC/pt2CCSD are in broad mutual agreement for this system.

\section{Conclusion}
\label{sec:conclusion}

In this work, we develop two size-extensive Auxiliary-Field Quantum Monte Carlo methods, AFQMC/ptCCSD and AFQMC/pt2CCSD, which maintain the same computational scaling as AFQMC/CISD. As demonstrated by the numerical examinations of non-interacting monomers and one-dimensional hydrogen chains, the lack of size-extensivity severely restricts the application of otherwise highly accurate methods, such as AFQMC/CISD, to extended systems. In contrast, the proposed PT and PT2 methods significantly outperform AFQMC/CISD in these regimes, with comparable magnitudes and scaling of stochastic noise.

Through a comprehensive set of benchmarks from main-group molecules to 3d transition metal complexes, AFQMC/pt2CCSD consistently yields more accurate results than AFQMC/ptCCSD, suggesting that incorporating the Coupled-Cluster $e^{\hat{T}_1}$ exactly is essential. Notably, we find little difference between AFQMC/pt2CCSD and AFQMC/stoCCSD, which stochastically samples the difference between $e^{T_2}$ and $1+ T_2$ so that the full CCSD trial is recovered without bias. This equivalence indicates that the perturbative treatment of AFQMC/pt2CCSD introduces negligible error, and within the framework of the CCSD trial wavefunction, there is no need to go beyond the PT2 level to capture additional correlation. 

Compared to CCSD(T) — the gold standard of quantum chemistry — AFQMC/pt2CCSD demonstrates superior accuracy in all systems examined in this work except for the diatomic transition metal monoxides. The degraded performance in that case may be attributed to the use of a Hartree-Fock guiding wavefunction in the phaseless approximation, which governs the force bias and the walkers' weights. This defect may be mitigated in future work by developing improved guiding wavefunctions or phaseless approximation schemes that preserve size-extensivity.

Additionally, a major merit of size-extensivity is its natural compatibility with local correlation techniques \cite{li2002linear,  rolik2011general, rolik2013efficient, nagy2019approaching}. Because size-extensivity ensures that the sum of fragment energies exactly equals the total energy of the macroscopic system, it facilitates fragmentation approaches that map computationally expensive, large-scale calculations onto smaller, localized active spaces. Building on recent advancements that achieved linear scaling for AFQMC with a Hartree-Fock trial \cite{kurian2023linear}, a local correlation variant of AFQMC/pt2CCSD is currently under development in our group. 

The methodologies presented in this work offer significant promise for broader applications. Unlike CCSD(T), the PT/PT2 approaches do not suffer from the infrared divergences that plague systems with low-energy excitations, opening the door for them to be applied to metals and systems undergoing insulator-metal transitions. Finally, the perturbative approach developed in this work is conceptually general; it can be adapted for any correlated wavefunction utilizing an exponential ansatz, paving the way for further theoretical explorations and novel applications.

\section{Data Availability}
Mean-field and Coupled Cluster calculations in this work are performed with \textsc{PySCF}~\cite{sun2018pyscf, sun2020pyscf}, which can be installed following the instructions on \url{https://github.com/pyscf/pyscf.git}. The AFQMC interface to \textsc{PySCF} used in this work can be found at the author's GitHub: \url{https://github.com/zh-yichi/afqmc}. Scripts to reproduce the uniform electron gas calculations can be found at: \url{https://github.com/zh-yichi/ueg}. The Supporting Information containing numerical results, data analysis, and scripts required to reproduce every calculation in this work is available at: \url{https://github.com/zh-yichi/afqmc_ptccsd_data}. Stabilized features of AFQMC will be released at \textsc{TROT} on \url{https://github.com/ankit76/trot.git}, which is a part of \textsc{PySCF}-Forge.

\section{Acknowledgements}
Y.Z and Y.D were partially supported by NSF CHE-
2145209. S.S. was partly supported by the DOE grant DE-SC0025943. Y.Z. thanks Eirik Kj\o nstad for discussions. A.M. thanks David Reichman for support. The computations were run using the allocation provided through the ACCESS project CHE240162.

\bibliography{reference}

\clearpage

\setcounter{section}{0}
\renewcommand{\thesection}{S\arabic{section}}
\setcounter{figure}{0}
\renewcommand{\thefigure}{S\arabic{figure}}
\setcounter{table}{0}
\renewcommand{\thetable}{S\arabic{table}}

\begin{center}
{\large\textbf{Supporting Information}}\\[4pt]
Size Extensive Auxiliary-Field Quantum Monte Carlo with\\
Perturbative Coupled Cluster Trial Wavefunction
\end{center}

\section{HEAT Dataset: Ground-State and Atomization Energies}

The HEAT (High-Accuracy Extrapolated Ab Initio Thermochemistry) dataset
comprises 26 small main-group molecules. We compute ground-state and
atomization energies with the cc-pVDZ basis set.  The CCSDTQP energies of
Bomble et al.\ serve as the reference, except for CO$_2$ for which the
reference is taken from Ref.~[12] of the main text.  All errors are reported
in mHa relative to the CCSDTQP reference; stochastic uncertainties are given
in parentheses.

\subsection{Per-molecule ground-state energy errors}

\begin{figure}[h!]
  \centering
  \includegraphics[width=\linewidth]{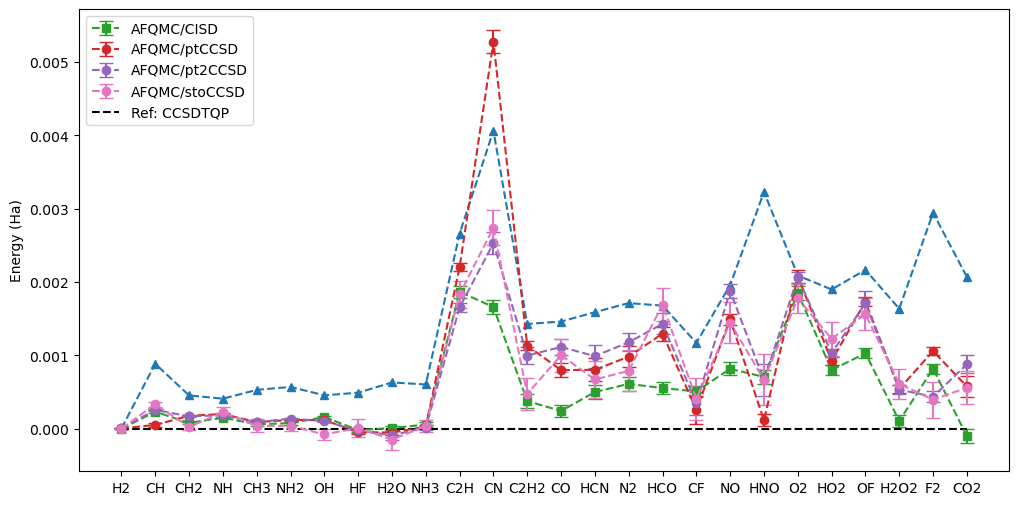}
  \caption{Ground-state energy error (mHa) relative to CCSDTQP for the 26
    molecules in the HEAT dataset (cc-pVDZ basis).  Error bars indicate the
    stochastic uncertainty (one standard deviation).}
  \label{fig:heat_gs}
\end{figure}

Figure~\ref{fig:heat_gs} shows the error in the ground-state energy of each
molecule relative to the CCSDTQP reference.  AFQMC/CISD achieves the lowest
overall root-mean-square deviation (RMSD = 0.75 mHa).  AFQMC/pt2CCSD
(RMSD = 1.1 mHa) slightly exceeds AFQMC/CISD but improves upon
AFQMC/ptCCSD (RMSD = 1.4 mHa) and CCSD(T) (RMSD = 1.79 mHa).
AFQMC/pt2CCSD and AFQMC/stoCCSD yield statistically identical results
throughout the dataset, confirming that the perturbative treatment of
$e^{\hat{T}_2}$ introduces negligible error.  The largest outlier for PT
relative to PT2 is CN, where the exact treatment of $e^{\hat{T}_1}$ in PT2
provides a marked improvement.

\subsection{Per-molecule atomization energy errors}

\begin{figure}[h!]
  \centering
  \includegraphics[width=\linewidth]{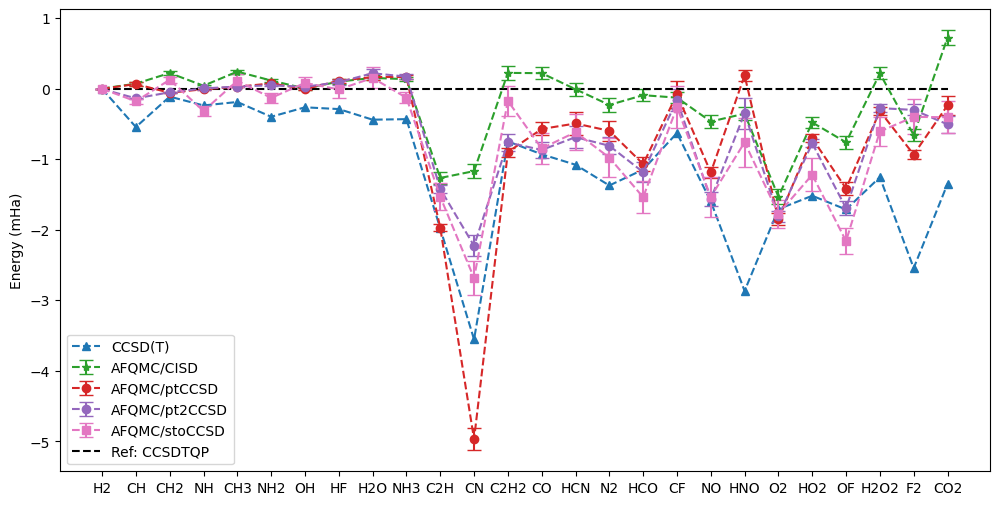}
  \caption{Atomization energy error (mHa) relative to CCSDTQP for the 26
    molecules in the HEAT dataset (cc-pVDZ basis).}
  \label{fig:heat_atom}
\end{figure}

Figure~\ref{fig:heat_atom} shows the error in the atomization energy,
$\Delta E = \sum_{a\in m} E_a - E_m$.  The RMSD values are summarized in
Table~III of the main text.  The ranking of methods is consistent with the
ground-state energies.  AFQMC/pt2CCSD (RMSD = 0.9 mHa) outperforms CCSD(T)
(RMSD = 1.42 mHa) and remains within chemical accuracy relative to the
reference.

\section{W4-MR Dataset: Ground-State and Atomization Energies}

\begin{sloppypar}
The W4-MR (Weizmann-4 Multi-Reference) dataset consists of 17 larger,
strongly correlated molecules with significant multi-reference character,
generated via the Weizmann-4 thermochemistry protocol.  We compute
ground-state and atomization energies with the cc-pVDZ basis.  CCSDTQ
energies serve as the reference for all molecules except ClO$_3$, for which
the AFQMC/HCI result is used.  The larger improvement of PT2 over PT
compared to the HEAT dataset highlights the importance of an exact treatment
of $e^{\hat{T}_1}$ for strongly correlated systems.
\end{sloppypar}

\subsection{Per-molecule ground-state energy errors}

\begin{figure}[h!]
  \centering
  \includegraphics[width=\linewidth]{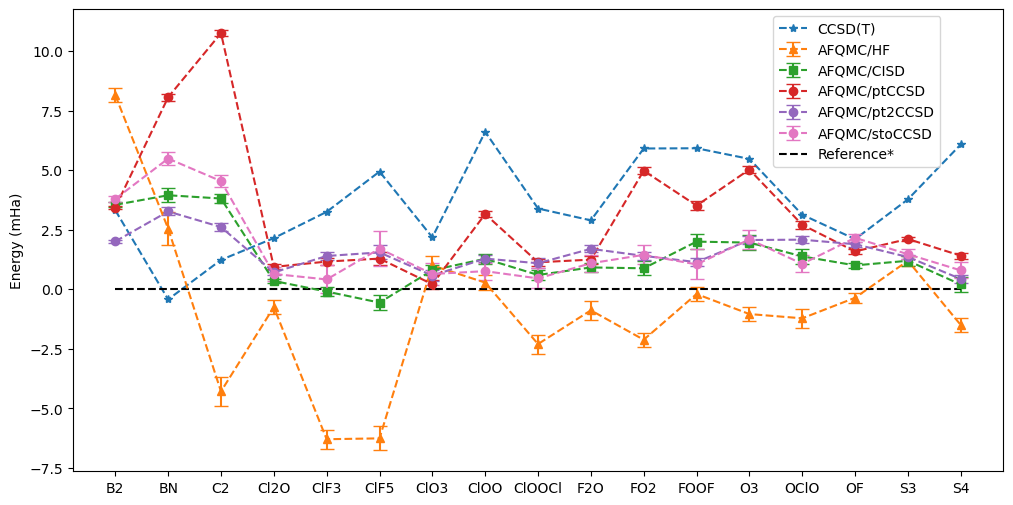}
  \caption{Ground-state energy error (mHa) relative to the reference
    (CCSDTQ or AFQMC/HCI for ClO$_3$) for the 17 molecules in the W4-MR
    dataset (cc-pVDZ basis).}
  \label{fig:w4mr_gs}
\end{figure}

Figure~\ref{fig:w4mr_gs} shows the ground-state energy error per molecule.
Notably, AFQMC/pt2CCSD (RMSD = 1.7 mHa) achieves accuracy comparable to
AFQMC/CISD (RMSD = 1.9 mHa), and both methods substantially outperform
CCSD(T) (RMSD = 4.1 mHa).  The large improvement from PT (RMSD = 4.1 mHa)
to PT2 is mainly attributable to strongly correlated molecules such as
BN, C$_2$, and O$_3$, where the singles amplitudes $T_1$ are large.

\subsection{Per-molecule atomization energy errors}

\begin{figure}[h!]
  \centering
  \includegraphics[width=\linewidth]{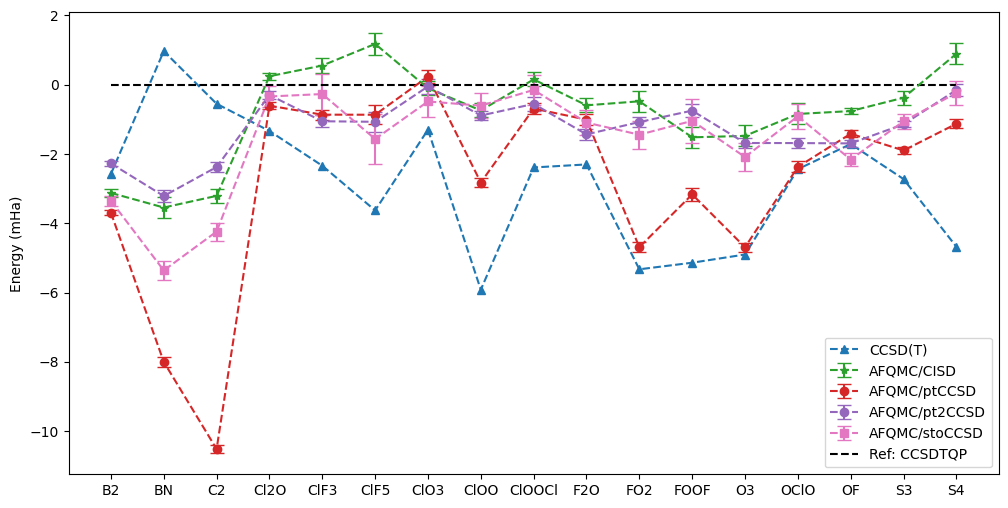}
  \caption{Atomization energy error (mHa) relative to the reference for the
    17 molecules in the W4-MR dataset (cc-pVDZ basis).}
  \label{fig:w4mr_atom}
\end{figure}

Figure~\ref{fig:w4mr_atom} shows the atomization energy error per molecule.
AFQMC/pt2CCSD (RMSD = 1.5 mHa) is comparable to AFQMC/CISD (RMSD = 1.6 mHa)
and substantially outperforms CCSD(T) (RMSD = 3.37 mHa).

\section{N$_2$: Convergence with Number of Stochastic Samples}

The AFQMC/stoCCSD method stochastically samples the difference
$e^{\hat{T}_2} - 1 - \hat{T}_2$ via the Hubbard--Stratonovich transformation,
using $N_s$ auxiliary-field samples per walker per time step.  A key question
is how many samples $N_y$ are needed to converge the energy without bias.

\begin{figure}[h!]
  \centering
  \includegraphics[width=0.85\linewidth]{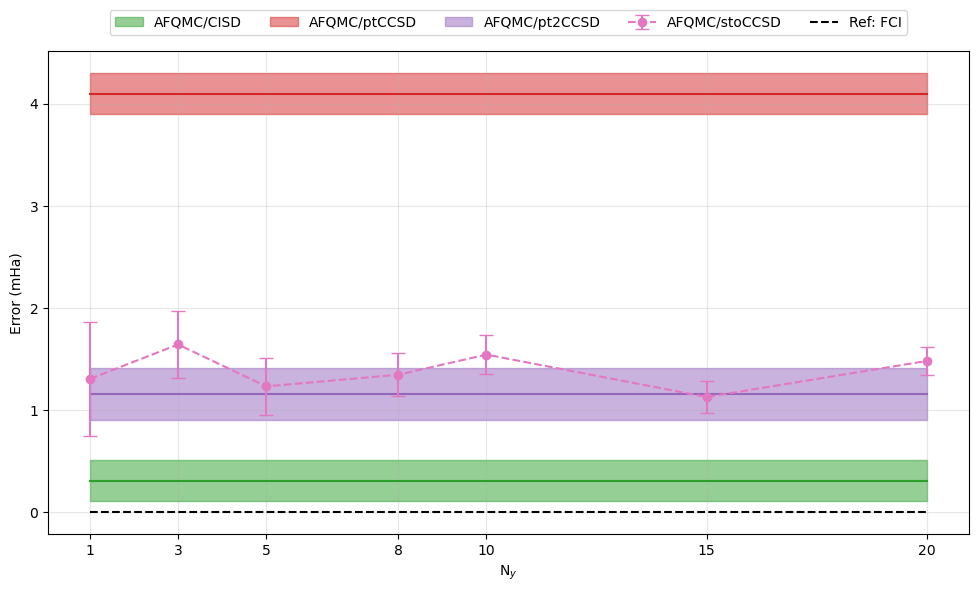}
  \caption{Ground-state energy error (mHa relative to FCI) of N$_2$ 
    at bond length = 3.6 Borh. AFQMC/stoCCSD energies are obtained with $N_y$ = 1, 3, 5, 8, 10, 15, and 20 stochastic trial samples per walker. The FCI reference energies are taken from Chan et al. All calculations use cc-pVDZ basis, UHF orbitals}
  \label{fig:n2_ny}
\end{figure}

Figure~\ref{fig:n2_ny} shows the AFQMC/stoCCSD energy for N$_2$ with bond length = 3.6 Bohr (cc-pVDZ basis, spin-unrestricted orbitals). The number of stochastic samples used to apply the CCSD trial $N_y$ = 1, 3, 5, 8, 10, 15, and 20.  Despite a reduction in the stochastic noise, the AFQMC/stoCCSD energies show no systematic dependence on $N_y$ in Figure~\ref{fig:n2_ny}, demonstrating that a single sample of $e^{\hat{T}_2} - 1 - \hat{T}_2$ per walker is
sufficient to faithfully represent the CCSD trial wavefunction without
introducing a systematic bias.  The perturbative PT2 results lie within
stochastic error of the stoCCSD results at all bond lengths, further
validating the perturbative approximation used in AFQMC/pt2CCSD.

\section{3d Transition Metal Monoxides: Dissociation Energies}

We benchmark AFQMC/ptCCSD and AFQMC/pt2CCSD on the seven $3d$ transition
metal monoxides ScO, TiO, VO, CrO, MnO, FeO, and CuO using the
Trial-Needs pseudopotential and corresponding basis set.  The dissociation
energy is extrapolated to the complete basis-set (CBS) limit via an inverse
cubic fit of the TZ and QZ energies.  SHCI energies serve as the reference.
All calculations are performed in spin-unrestricted orbitals.

\begin{figure}[h!]
  \centering
  \includegraphics[width=\linewidth]{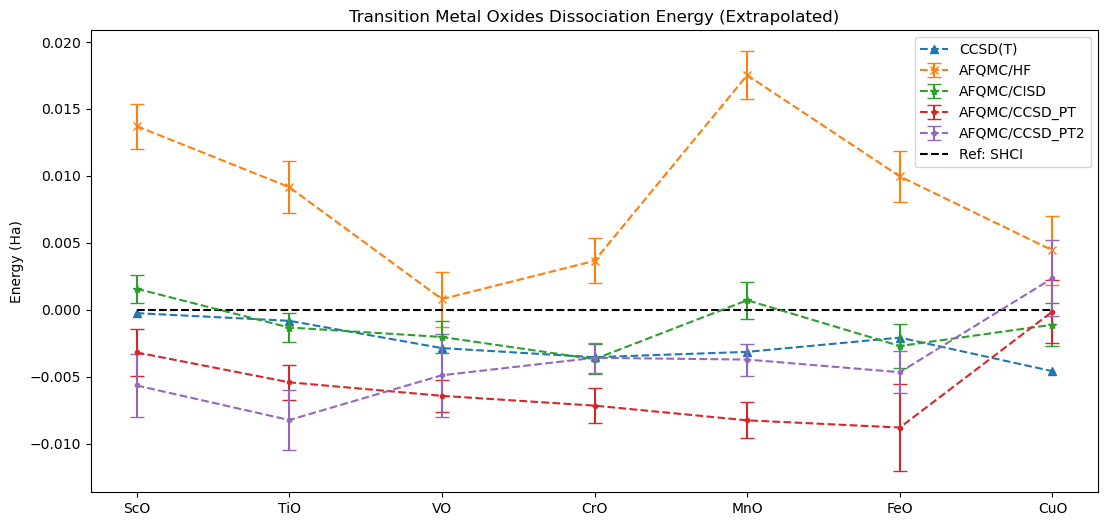}
  \caption{CBS-extrapolated dissociation energy error (mHa) relative to SHCI
    for the seven $3d$ transition metal monoxides.  The CBS limit is obtained
    via an inverse cubic fit of the TZ and QZ results.}
  \label{fig:tmo_diss}
\end{figure}

Figure~\ref{fig:tmo_diss} shows the CBS-extrapolated dissociation (atomization)
energy error for each monoxide.  The RMSD values are summarized in Table~IX
of the main text.  AFQMC/pt2CCSD (RMSD $\approx 5$ mHa) performs similarly to
AFQMC/ptCCSD and somewhat worse than AFQMC/CISD (RMSD $\approx 2$ mHa),
while all three AFQMC methods are competitive with or superior to CCSD(T)
(RMSD $\approx 2.85$ mHa) for most individual systems.  We suspect that the use of the
Hartree--Fock guiding wavefunction is the primary source of error for these
systems; future work on improved guiding wavefunctions or size-extensive
phaseless approximations is expected to close this gap.

\end{document}